\documentclass[twocolumn,tighten]{aastex631}

\usepackage{newtxtext,newtxmath}
\usepackage{natbib}
\RequirePackage{fix-cm}

\usepackage{soul}

\newcommand{\del}[1]{\textcolor{magenta}{{\iffalse{#1}\fi}}}

\begin{document}

\title{
The FAST Galactic Plane Pulsar Snapshot Survey: VII. Six millisecond pulsars in compact orbits with massive white dwarf companions}

\correspondingauthor{J.~L. Han}

\author[0009-0009-6590-1540]{Z.~L. Yang}
\affiliation{National Astronomical Observatories, Chinese Academy of Sciences, Jia-20 Datun Road, 
ChaoYang District, Beijing
   100012, China}
\affiliation{School of Astronomy and Space Science, University of Chinese Academy of Sciences, Beijing 100049, China}

\author[0000-0002-9274-3092]{J.~L. Han}\thanks{E-mail: hjl@nao.cas.cn}
\affiliation{National Astronomical Observatories, Chinese Academy of Sciences, Jia-20 Datun Road, 
ChaoYang District, Beijing
   100012, China}
\affiliation{School of Astronomy and Space Science, University of Chinese Academy of Sciences, Beijing 100049, China}
\affiliation{Key Laboratory of Radio Astronomy and Technology,  Chinese Academy of Sciences, Beijing 100101, China }

\author{T. Wang}
\affiliation{National Astronomical Observatories, Chinese Academy of Sciences, Jia-20 Datun Road, 
ChaoYang District, Beijing
   100012, China}

\author[0000-0002-6437-0487]{P.~F. Wang}
\affiliation{National Astronomical Observatories, Chinese Academy of Sciences, Jia-20 Datun Road, 
ChaoYang District, Beijing
   100012, China}
\affiliation{School of Astronomy and Space Science, University of Chinese Academy of Sciences, Beijing 100049, China}
\affiliation{Key Laboratory of Radio Astronomy and Technology,  Chinese Academy of Sciences, Beijing 100101, China }

\author[0009-0003-2212-4792]{W.~Q. Su}
\affiliation{National Astronomical Observatories, Chinese Academy of Sciences, Jia-20 Datun Road, ChaoYang District, Beijing 100012, China}
\affiliation{School of Astronomy and Space Science, University of Chinese Academy of Sciences, Beijing 100049, China}

\author{W.~C. Chen}
\affiliation{School of Science, Qingdao University of Technology, Qingdao 266525, China}

\author[0000-0002-9274-3092]{C. Wang}
\affiliation{National Astronomical Observatories, Chinese Academy of Sciences, Jia-20 Datun Road, 
ChaoYang District, Beijing
   100012, China}
\affiliation{School of Astronomy and Space Science, University of Chinese Academy of Sciences, Beijing 100049, China}
\affiliation{Key Laboratory of Radio Astronomy and Technology,  Chinese Academy of Sciences, Beijing 100101, China }
   
\author[0000-0002-6423-6106]{D.~J. Zhou}
\affiliation{National Astronomical Observatories, Chinese Academy of Sciences, Jia-20 Datun Road, 
ChaoYang District, Beijing
   100012, China}

\author[0009-0008-1612-9948]{Y. Yan}
\affiliation{National Astronomical Observatories, Chinese Academy of Sciences, Jia-20 Datun Road, 
ChaoYang District, Beijing
   100012, China}
\affiliation{School of Astronomy and Space Science, University of Chinese Academy of Sciences, Beijing 100049, China}

\author[0000-0002-1056-5895]{W.~C. Jing}
\affiliation{National Astronomical Observatories, Chinese Academy of Sciences, Jia-20 Datun Road, 
ChaoYang District, Beijing
   100012, China}
\affiliation{School of Astronomy and Space Science, University of Chinese Academy of Sciences, Beijing 100049, China}

\author{N.~N. Cai}
\affiliation{National Astronomical Observatories, Chinese Academy of Sciences, Jia-20 Datun Road, 
ChaoYang District, Beijing
   100012, China}

\author{L. Xie}
\affiliation{National Astronomical Observatories, Chinese Academy of Sciences, Jia-20 Datun Road, 
ChaoYang District, Beijing
   100012, China}

\author{J. Xu}
\affiliation{National Astronomical Observatories, Chinese Academy of Sciences, Jia-20 Datun Road, 
ChaoYang District, Beijing
   100012, China}
\affiliation{Key Laboratory of Radio Astronomy and Technology,  Chinese Academy of Sciences, Beijing 100101, China }

\author{H.~G. Wang}
\affiliation{Department of Astronomy, School of Physics and Materials Science, Guangzhou University, Guangzhou 510006, Guangdong Province, China }
\affiliation{National Astronomical Data Center, Great Bay Area, Guangzhou 510006, Guangdong Province, China }

\author{R.~X. Xu}
\affiliation{Department of Astronomy, Peking University, Beijing 100871, China }




\begin{abstract}
Binary millisecond pulsars with a massive white dwarf (WD) companion are intermediate-mass binary pulsars (IMBPs). They are formed via the Case BB Roche-lobe overflow evolution channel if they are in compact orbits with an orbital period of less than 1 day. They are fairly rare in the known pulsar population; only five such IMBPs have been discovered before, and one of them is in a globular cluster. Here we report six IMBPs in compact orbits: PSRs J0416+5201, J0520+3722, J1919+1341, J1943+2210, J1947+2304 and J2023+2853, discovered during the Galactic Plane Pulsar Snapshot survey by using the Five-hundred-meter Aperture Spherical radio Telescope, doubling the number of such IMBPs due to the high survey sensitivity in the short survey time of 5 minutes. Follow-up timing observations show that they all have either a CO WD or an ONeMg WD companion with a mass greater than about 0.8~$M_\odot$ in a very circular orbit with an eccentricity in the order of $\lesssim10^{-5}$. PSR J0416+5201 should be an ONeMg WD companion with a remarkable minimum mass of 1.28 $M_\odot$. These massive WD companions lead to a detectable Shapiro delay for PSRs J0416+5201, J0520+3722, J1943+2210, and J2023 +2853, indicating that their orbits are highly inclined. From the measurement of the Shapiro delay, the pulsar mass of J1943+2210 was constrained to be 1.84$^{\,+0.11}_{-0.09}$~$M_\odot$, and that of PSR J2023+2853 to be 1.28$^{\,+0.06}_{-0.05}$~$M_\odot$. 
\end{abstract}

\keywords{Pulsars: general - binaries: close - stars: evolution}

\section{Introduction} \label{sec:intro}

\begin{figure}[t]
    \centering
    \includegraphics[width=0.98\columnwidth]{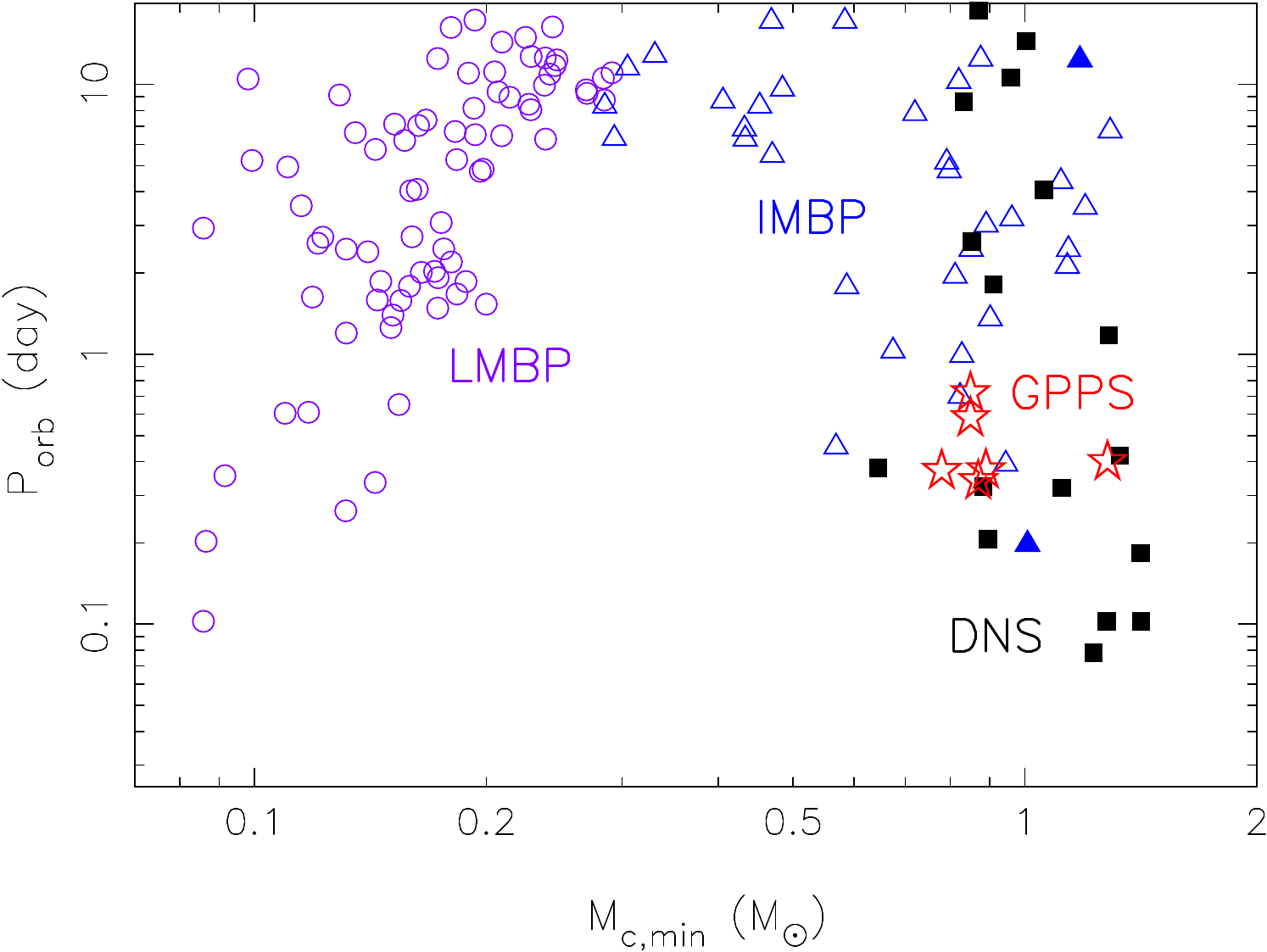}
    \caption{Six intermediate-mass binary pulsars (IMBP, stars) in this paper are plotted among pulsars with a degenerate companion in the distribution of orbital period and minimum companion mass. The companions of low-mass binary pulsars (LMBP, circles) are He white dwarfs, while the intermediate-mass binary millisecond pulsars (IMBP, triangles) have a CO/ONeMg WD companion. Pulsars in double neutron star (DNS) binary systems are marked by black  filled squares. Binaries with an orbital eccentricity greater than 0.01 are marked with filled symbols. The minimum companion mass is derived from the mass function, assuming a pulsar mass of 1.4 $M_\odot$. Parameters are taken from the ATNF pulsar catalog  \citep{Manchester+2005AJ....129.1993M} (web version on 2024 September 27th), except for six IMBPs from this paper. }
    \label{pb-m2}
\end{figure}

Binary millisecond pulsars (MSPs) with a CO/ONeMg white dwarf (WD) companion are intermediate-mass binary pulsars (IMBPs), because they have a more massive companion (see Figure~\ref{pb-m2}) compared to these low-mass binary pulsars with a He WD companion with a mass less than 0.46 $M_\odot$\citep{Tauris+2023pbse.book.....T}. 
When the WD companion has a mass of $\gtrsim1.05$ $M_\odot$, it must be an ONeMg WD. The CO WDs have a mass roughly in the range of 0.46 -- 1.05  $M_\odot$. 
The IMBP systems are descendants of intermediate-mass X-ray binaries (IMXBs) where mass is transferred from a more massive donor to a less massive neutron star (NS). The most widely accepted explanation for the high spin frequencies of MSPs is the accretion of mass and angular momentum in a binary system \citep{Alpar+1982Natur.300..728A,  Radhakrishnan+1982CSci...51.1096R, Bhattacharya+1991PhR...203....1B}. The spin periods are therefore closely related to the amount of accreted mass in this process and, hence, provide constraints on the accretion process \citep{Tauris+2011MNRAS.416.2130T, Tauris+2012MNRAS.425.1601T}. The mass transfer process can be dynamically unstable, leading to the formation of a common envelop (CE), which experiences much more complicated and unclear evolution depending on the mass of the donor star \citep{Paczynski+1976IAUS...73...75P, Iben+1993PASP..105.1373I, Ivanova+2013A&ARv..21...59I, Tauris+2023pbse.book.....T}. During the H-shell burning phase, the donor star with a mass of 2 -- 10 $M_\odot$ expands and enters the red giant branch. At the start of mass transfer, if the donor star is near the tip of the red giant branch and the merger can be avoided, an unevolved helium star will be left and then the post-CE orbital period will be $\leq$ 0.5 days \citep{Tauris+2012MNRAS.425.1601T}. The helium star will expand during the helium shell burning phase, leading to a new mass transfer called the Case BB Roche-lobe overflow \citep[RLO;][]{Tauris+2011MNRAS.416.2130T, Tauris+2012MNRAS.425.1601T, Tauris+2023pbse.book.....T}. An IMBP system can then be formed with a very tight orbit. During these mass transfer phases, the NS accretes only a little mass from its companion and is only mildly recycled.

Timing observations of binary MSPs provide an effective approach to studying IMBPs. The Keplerian binary parameters can be determined with a few observations \citep{Freire+2001MNRAS.322..885F, Bhattacharyya+2008MNRAS.387..273B}, then the companion mass can be estimated based on the mass function. If any post- Keplerian (PK) binary parameters are precisely measured, then the masses of the pulsar and the WD companion can be better constrained under the general relativity frame \citep{Damour+1992PhRvD..45.1840D}. The proper motions, if measurable from long-term timing, can tell us the natal kick of the pulsar obtained during the formation of the binary systems.

Using the Five-hundred-meter Aperture Spherical radio Telescope \citep[FAST,][]{Nan2006,2011IJMPD..20..989N} which is the most sensitive single-dish radio telescope in the world, we are carrying out the FAST Galactic Plane Pulsar Snapshot (GPPS) survey aiming at discovering pulsars within the Galactic latitude of $\pm 10^\circ$ of the FAST visible sky area \citep{Han+2021RAA....21..107H}. The survey observations are made with 4 paintings to cover a sky patch of 0.1575 square degrees, with an integration time of each pointing for 300~s, which gives the sensitivity of a few $\mu$Jy for pulsars in millisecond range, improved by two magnitudes compared to previous surveys \citep[see Figure6 in][]{Han+2021RAA....21..107H}. The FAST GPPS survey has already successful discovered more than 750 new pulsars\footnote{http://zmtt.bao.ac.cn/GPPS/GPPSnewPSR.html} \citep{Han+2024RAA}. Among them, there are more than 170 millisecond pulsars and about 116 pulsars are in binary systems \citep{Wang+2024RAA}. The high sensitivity of FAST in such a short integration time is good at detecting pulsars in compact orbits. Here we report the timing results of six binary pulsars: PSRs J0416+5201, J0520+3722, J1919+1341, J1943+2210, J1947+2304, and J2023+2853, with a companion more massive than 0.8 $M_\odot$. Because their orbits are nearly circular with a period of less than 1 day and ellipticity on the order of $10^{-5}$, they must be IMBPs with either a CO WD or an ONeMg WD, with the parameter space mixed with double NSs (see Figure~\ref{pb-m2}). The new discovery of six pulsars by the GPPS survey significantly enlarges the number of IMBPs in compact orbits (see Table~\ref{compact IMBP}). 

The structure of the rest of this paper is arranged as follows: In Section 2, we describe the FAST observations of these pulsars, followed by the details of data analyses including the orbital parameter determination and pulsar timing. The detailed results of these six pulsars are presented in Section 3. The implications of our results and future prospects are discussed in Section 4.

\begin{table}
\centering
	\caption{Intermediate-mass Binary Millisecond Pulsars in Compact Circular Orbits of $P_{\rm orb} < 1$ day and $e<0.01$, Six are from this Paper and Five are Previously Known
 }
	\label{compact IMBP}
	\setlength{\tabcolsep}{2mm}{
	\renewcommand\arraystretch{1.0}
    {
    \begin{tabular}{lccccc}
    \hline
   PSR-name & $P$  & $P_{\rm orb}$ & $x$ & $M_{\rm c,min}$  & Ref.  \\
            & (ms) & (days)        & (ls) & ($M_\odot$)  & \\
    \hline
    J1947+2304 & 10.89 & 0.339 & 2.40 & 0.87 & [0] \\
    J1919+1341 & 11.66 & 0.370 & 2.34 & 0.78 & [0] \\
    J1943+2210 & 12.87 & 0.372 & 2.58 & 0.89 & [0] \\
    J1748$-$2446N & 8.667 & 0.386 & 1.62 & 0.48 & [1] \\
    J1952+2630 & 20.73 & 0.392 & 2.80 & 0.94 &  [2] \\
    J0416+5201 & 18.24 & 0.396 & 3.51 & 1.28 & [0] \\
    J1757$-$5322 & 8.870 & 0.453 & 2.09 & 0.57 & [3]  \\
    J0520+3722 & 7.913 & 0.580 & 3.37 & 0.85 & [0] \\
    J1802$-$2124 & 12.65 & 0.699 & 3.72 & 0.82 & [4]  \\
    J2023+2853 & 11.33 & 0.718 & 4.00 & 0.89 & [0] \\ 
    J1525$-$5545 & 11.36 & 0.990 & 4.71 & 0.83 & [5] \\
    \hline
    \end{tabular}}\\
    References: 
    [0]: this paper; 
    [1]: \citet{J1748-2446N+Ransom2005}; 
    [2]: \citet{J1952+2630+Knispel2011}; 
    [3]: \citet{J1757-5322+Edwards2001};
    [4]: \citet{J1802-2124+Faulkner2004}; 
    [5]: \citet{J1525-5545+Ng2014}. 
       }   
\end{table}

\begin{figure*} 
    \centering
    \includegraphics[width=0.49\textwidth]{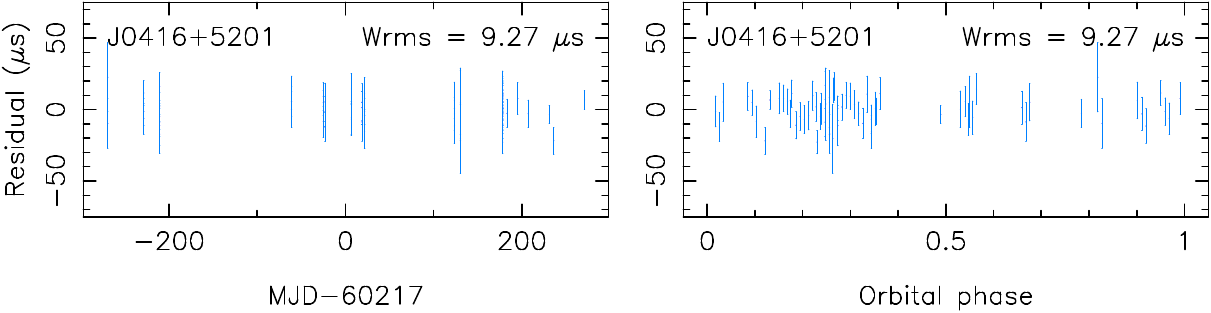} 
    \includegraphics[width=0.49\textwidth]{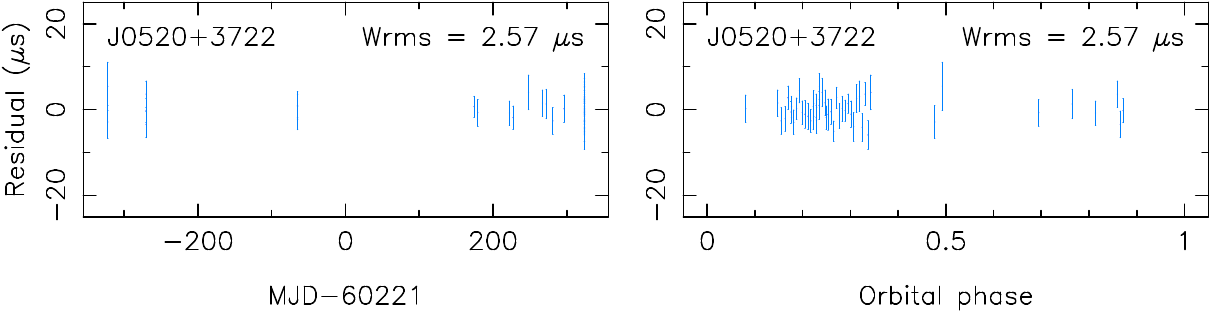} \\[1mm]
    \includegraphics[width=0.49\textwidth]{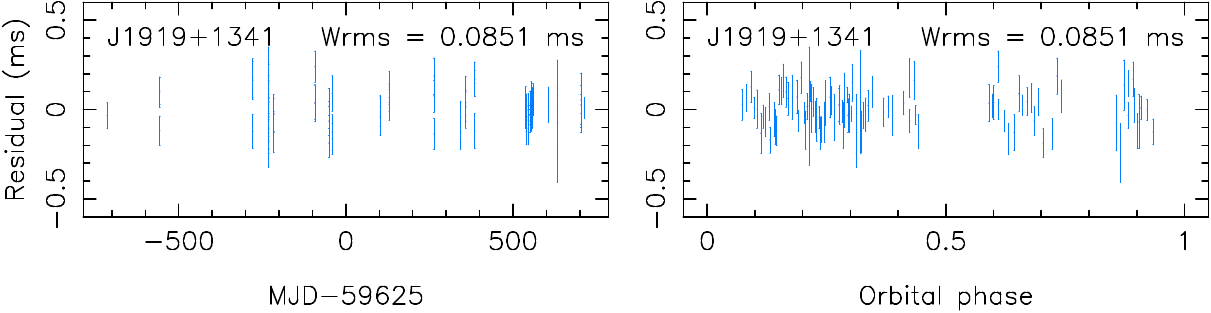} 
    \includegraphics[width=0.49\textwidth]{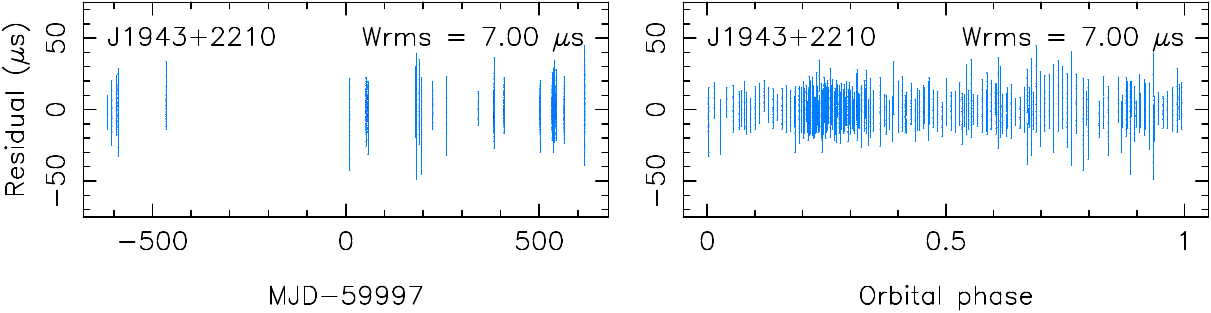} \\[1mm]
    \includegraphics[width=0.49\textwidth]{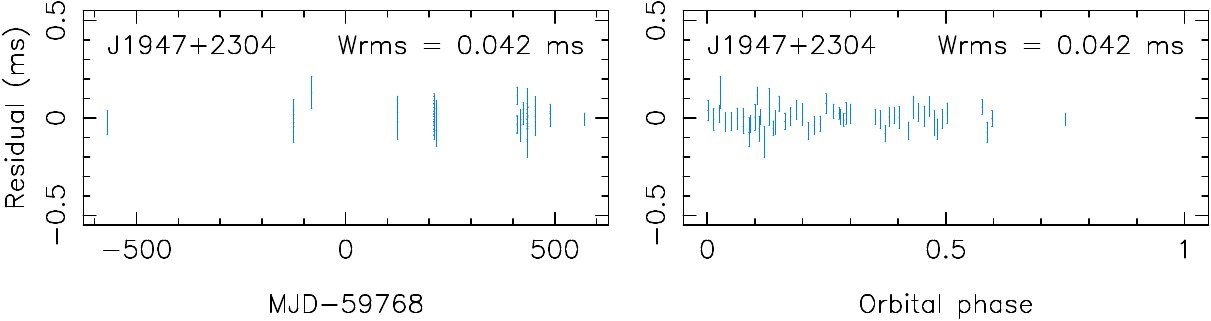} 
    \includegraphics[width=0.49\textwidth]{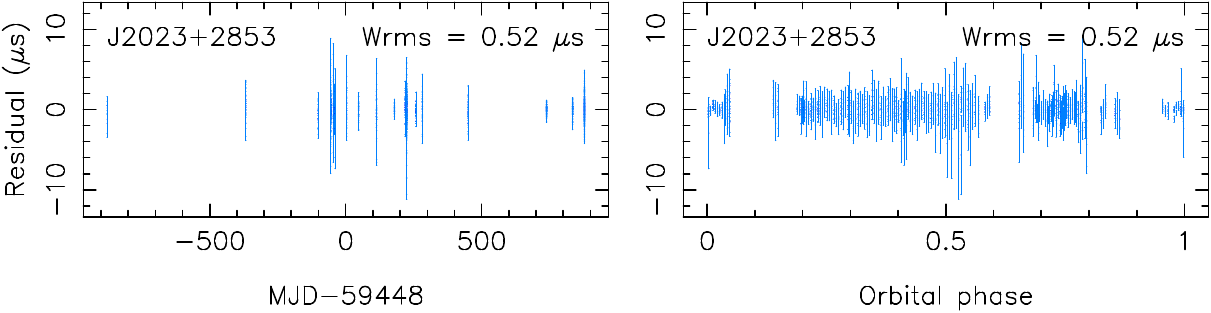} 
    \caption{Timing residuals of the 6 pulsars measured by FAST, plotted along observation date in MJD and against the orbital phases. The difference between measured TOAs and the derived timing model, $W_{\rm rms}$, are marked in each panel.
    }
    \label{timing_residuals_other}
\end{figure*}

\begin{table*}[t]
	\centering
\small
	\caption{Parameters of PSRs J0416+5201, J0520+3722 and J1919+1341 derived from FAST observations. The timing results are obtained through the package {\sc TEMPO2} \citep{Hobbs+2006MNRAS.369..655H} with the ELL1k model, in units of TCB. The uncertainty of each TOA is scaled by T2EFAC to set the reduced $\rm\chi^2$ to 1. The uncertainty of the model parameters in brackets is the 68.3\% confidence level (1$\sigma$ uncertainty). The minimum companion mass is derived from the mass function assuming a pulsar mass of 1.4 $M_\odot$. The flux density at 1.25 GHz was calculated by using {\sc PSRFLUX} in {\sc PSRCHIVE} \citep{Hotan+2004PASA...21..302H} for the calibrated profiles. 
    }
    \vspace{1mm}
	\label{timingsolution1}
	\setlength{\tabcolsep}{2mm}{
	\renewcommand\arraystretch{0.9}
    {
    \begin{tabular}{lccc}
    		\hline
    		Parameter  & PSR  J0416+5201  & PSR J0520+3722 & PSR J1919+1341\\
                       & (GPPS 0560)     &  (GPPS 0538)    &  (GPPS 0215)  \\
    		\hline                                 
    Right ascension, $RA$ (J2000)   & 04:16:27.27754(19) & 05:20:13.56522(5) & 19:19:23.0125(7) \\
    Declination, $DEC$ (J2000)   & +52:01:25.774(4) & +37:22:09.573(7) & +13:41:09.29(1) \\       
    Galactic longitude, $l$ (deg)  & 151.61631 & 170.03911 & 48.26673\\
    Galactic latitude, $b$ (deg)   & 0.95628  & 0.15093 & 0.15517 \\
    Spin frequency, $\nu$ (Hz)  & 54.82288789105(3) & 126.370150421555(7) & 85.79476748198(7)\\
    Spin frequency derivative, $\Dot{\nu}$ (Hz $\rm s^{-1}$)  & $-$7.001(16)$\rm\times10^{-16}$ & $-$5.694(2)$\rm\times10^{-15}$  & $-$1.25(2)$\rm\times10^{-16}$ \\
    Reference epoch (MJD)  & 60300 & 60300 & 59393\\
    Spin period, $P$ (s)  & 0.018240556790575(9) & 0.0079132611353561(4) & 0.01165572247993(1) \\
    Spin period derivative, $\Dot{P}$ (s/s) & 2.329(5)$\rm\times10^{-19}$ & 3.5653(15)$\rm\times10^{-19}$ & 1.70(3)$\rm\times10^{-20}$\\   
    Characteristic age,  $\rm \tau_c$ (Gyr) & 1.24 & 0.352 & 10.9 \\
    Dispersion measure (DM) (pc $\rm cm^{-3}$)  & 140.497(3) & 88.9858(9)  & 394.56(1) \\
    DM Distance from NE2001 / YMW16 (kpc)  & 4.4 / 2.8 &  2.3 / 1.7 & 9.9 / 8.7 \\ 
    Faraday rotation measure, $RM$ (rad $\rm m^{-2}$) & $-$196(3) & 28(3) & -\\
    Flux density at 1.25 GHz ($\mu$Jy) & 39.3(2) & 128.0(4) & 23.4(4)\\
     \hline
    Orbital period, $P_{\rm orb}$ (days) & 0.39646949377(13) & 0.57967552341(16)  & 0.3703382666(5) \\
    Projected semi-major axis, $x$ (ls)  & 3.505563(8) & 3.372430(3) & 2.34094(2) \\
    Time of ascending node passage, $ T_{\rm ASC}$ (MJD) & 60317.56895079(5) &  60298.52239671(5) & 59392.5733017(8) \\
    Mass function, $f(m)$ ($ M_\odot$) & 0.2942636(17) & 0.1225580(3) & 0.100429(2) \\
    Minimum companion mass ($M_\odot$) & 1.28 & 0.85 & 0.78 \\ 
    First Laplace parameter, $\epsilon_1=e\sin\omega$  & 2.2(3)$\rm\times10^{-5}$ & 2(2)$\rm\times10^{-6}$ & 0.8(1.8)$\rm\times10^{-5}$   \\
    Second Laplace parameter,  $\epsilon_2=e\cos\omega$ & 3.99(17)$\rm\times10^{-5}$ & $-$1.0(1.3)$\rm\times10^{-6}$ & 0.7(1.2)$\rm\times10^{-5}$ \\ 
    Sine of inclination angle, $\sin i$ & 0.988(8) & 0.9980(14) & - \\
    Companion mass, $m_{\rm c}$ ($M_\odot$) & 1.7(5) & 0.70(13) & -\\
    \hline
    \end{tabular}
    }
}
\end{table*}
\begin{table*}[t]
	\centering
\small
	\caption{Same as Table~\ref{timingsolution1} but for PSRs J1943+2210, J1947+2304, and J2023+2853. The Bayesian analyses have been carried out to get the pulsar and companion masses and the orbit inclination angle for the PSRs J1943+2210 and J2023+2853 binary. 
    }
    \vspace{1mm}
	\label{timingsolution2}
	\setlength{\tabcolsep}{2mm}{
	\renewcommand\arraystretch{0.9}
    {
    \begin{tabular}{lccc} 
    		\hline
    		Parameter       & PSR J1943+2210 & PSR J1947+2304 & PSR J2023+2853 \\
                            &  (GPPS 0227)   &  (GPPS 0379)   &  (GPPS 0201)   \\
      
    		\hline                                 
    		
    Right ascension, $RA$ (J2000)    & 19:43:53.77751(4) & 19:47:28.938(1) &  20:23:21.063406(6) \\
    Declination, $DEC$ (J2000)       & +22:10:33.9676(8) & +23:04:15.30(1) &  +28:53:41.4521(1) \\
    Galactic longitude, $l$ (deg)  & 58.512922 & 59.702179 & 68.944528  \\
    Galactic latitude, $b$ (deg)  & $-$0.852878 & $-$1.120372 & $-$4.809745 \\
    Proper motion in RA direction (mas yr$^{-1}$)  & $-$4.8(0.3) & - & $-$3.30(5) \\
    Proper motion in DEC direction (mas yr$^{-1}$) & $-$5.6(0.5) & - & $-$8.10(8) \\
    Parallax (mas) & - & - & 1.2(0.3)\\             
    Spin frequency, $\nu$ (Hz)   & 77.699600217448(5) & 91.79624431818(6) & 88.2697836893946(9) \\
    Spin frequency derivative, $\Dot{\nu}$ (Hz $\rm s^{-1}$) & $-$3.4172(8)$\rm\times10^{-16}$ & $-$3.39(7)$\rm\times10^{-16}$ & $-$2.0898(1)$\rm\times10^{-16}$\\
    Reference epoch (MJD)  & 59800 & 59800 & 59262 \\
    Spin period, $P$ (s) & 0.0128700790892287(8) &  0.010893691865365(7) & 0.0113289050703785(1)\\
    Spin period derivative, $\Dot{P}$ (s/s) & 5.6602(13)$\rm\times10^{-20}$ & 4.03(8)$\rm\times10^{-20}$ & 2.6821(2)$\rm\times10^{-20}$\\ 
    Characteristic age,  $\rm \tau_c$ (Gyr)  &  3.61 & 4.29 & 6.70 \\
    Dispersion measure (DM) (pc $\rm cm^{-3}$) & 110.6643(10) & 321.00(1) &   22.75353(8)\\
    DM Distance from NE2001 / YMW16 (kpc) & 4.6 / 4.0 & 9.8 / 10.2 & 2.0 / 1.6\\ 
    DM derivative (pc $\rm cm^{-3}$ yr$^{-1}$) & 0.0013(6) & - & 0.00017(4)\\
    Faraday rotation measure, $RM$ (rad $\rm m^{-2}$) & $-$37(3) & - &$-$65(3)\\
    Flux density at 1.25 GHz ($\mu$Jy) & 25.0(1) & 14.7(2) & 662.5(2)\\ 
     \hline
    Orbital period, $P_{\rm orb}$ (days) & 0.37205272916(4) & 0.3388818158(7) & 0.71823041855(1) \\
    Projected semi-major axis, $x$ (ls)  & 2.5798019(9) & 2.39709(4) &  4.0022204(3) \\
    Time of ascending node passage, $ T_{\rm ASC}$ (MJD) & 59408.78530783(4) & 59979.0730606(5) & 59211.1878597(3) \\
    Mass function, $f(m)$ ($M_\odot$)  & 0.13317810(13) & 0.128777(7) & 0.13343143(3)\\
    Minimum companion mass, m$_{\rm c,min}$ ($M_\odot$) & 0.89 & 0.87 & 0.89 \\ 
    First Laplace parameter, $\epsilon_1=e\sin\omega$  & $-$1.2(0.5)$\rm\times10^{-6}$ & 0.9(2.2)$\rm\times10^{-5}$ & 1.113(6)$\rm\times10^{-5}$ \\
    Second Laplace parameter,  $\epsilon_2=e\cos\omega$ & $-$1.0(0.3)$\rm\times10^{-6}$ & $-$1.4(1.6)$\rm\times10^{-5}$ & 0.694(5)$\rm\times10^{-5}$\\
    Rate of change of $P_{\rm orb}$, $ \Dot{P}_{\rm orb}$ (10$^{-13}$ s/s) & $-$1.3(0.7) & - & $-$0.9(0.3)\\
    Rate of change of $x$, $\Dot{x}$ (10$^{-15}$ ls s$^{-1}$) & - & - & $-$3(3)\\
    Rate of periastron advance, $\Dot\omega$ & - & - & 0.7(0.2)\\
    Sine of inclination angle, $\sin i$ & 0.99979(5) & - & 0.9939(4)\\
    Companion mass, $m_{\rm c}$ ($M_\odot$) & 1.03(3) & - & 0.85(2)\\
    \hline
    Pulsar mass ($M_\odot$) & 1.84$^{\,+0.11}_{-0.09}$ & - & 1.28$^{\,+0.06}_{-0.05}$\\ [1mm]
    Companion mass ($M_\odot$) & 1.03$^{\,+0.04}_{-0.03}$ & - & 0.85$^{\,+0.02}_{-0.02}$\\ [1mm]
    Orbital inclination angle (deg) & 88.80$^{\,+0.13}_{-0.14}$ & - &  83.7$^{\,+0.2}_{-0.3}$\\ [1mm]
    \hline
    \end{tabular}}
    \\
}
\end{table*}

\begin{figure*}[htp]
    \centering
    \includegraphics[width=0.32\textwidth]{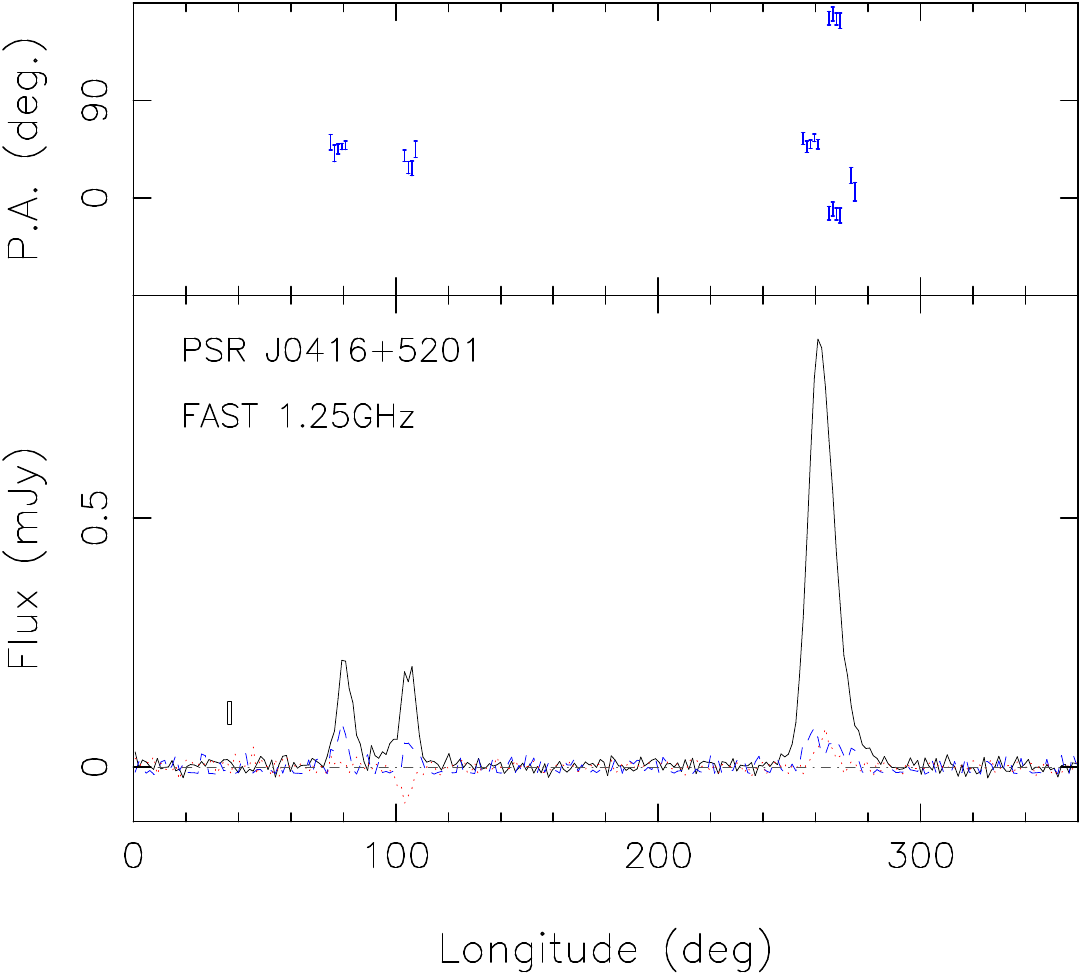}
    \includegraphics[width=0.32\textwidth]{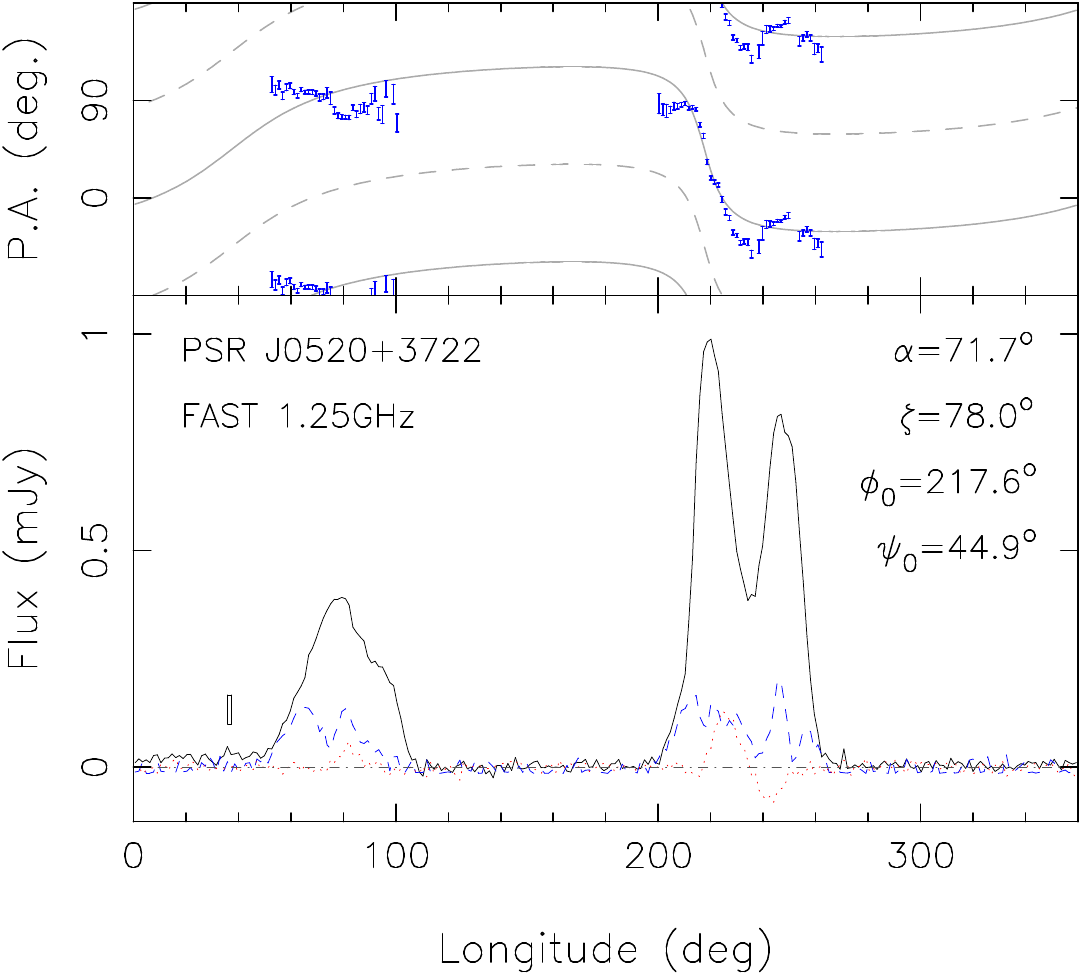}
    \includegraphics[width=0.32\textwidth]{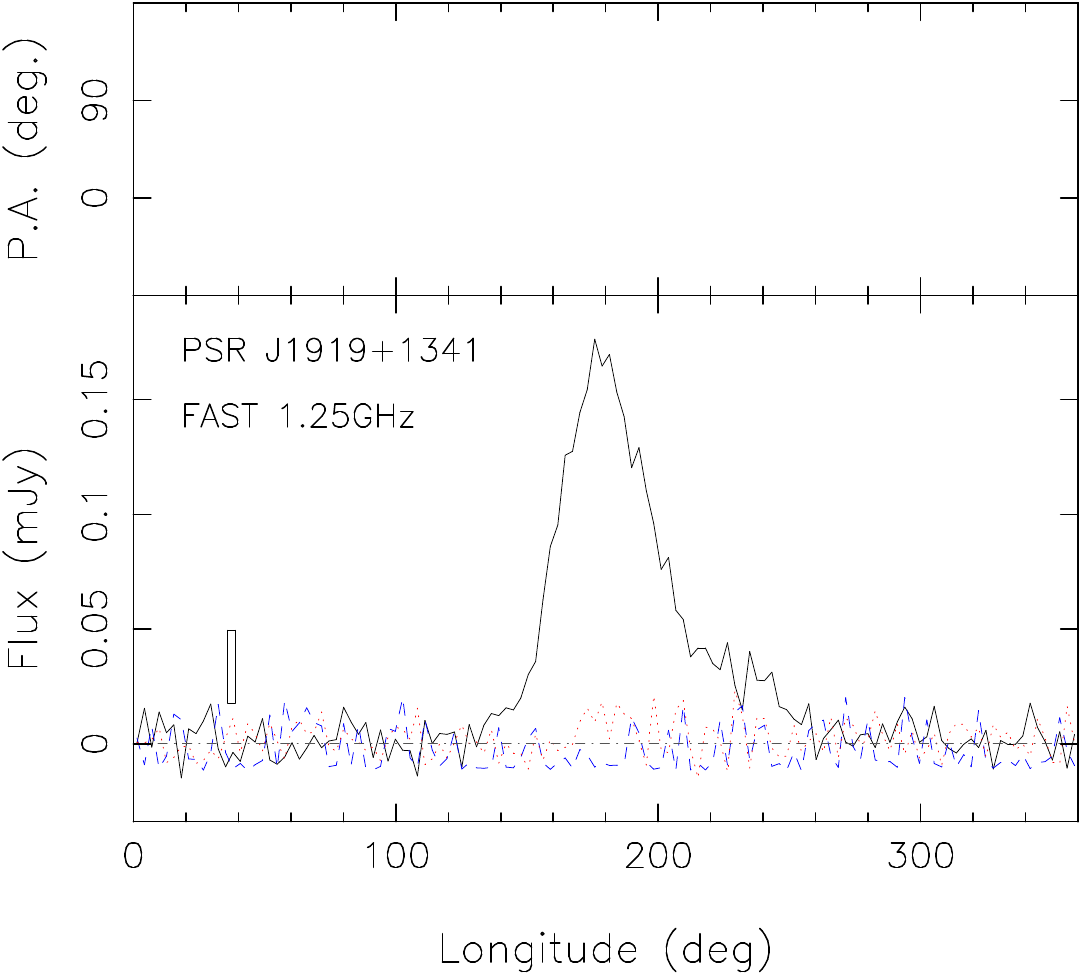}
    \includegraphics[width=0.32\textwidth]{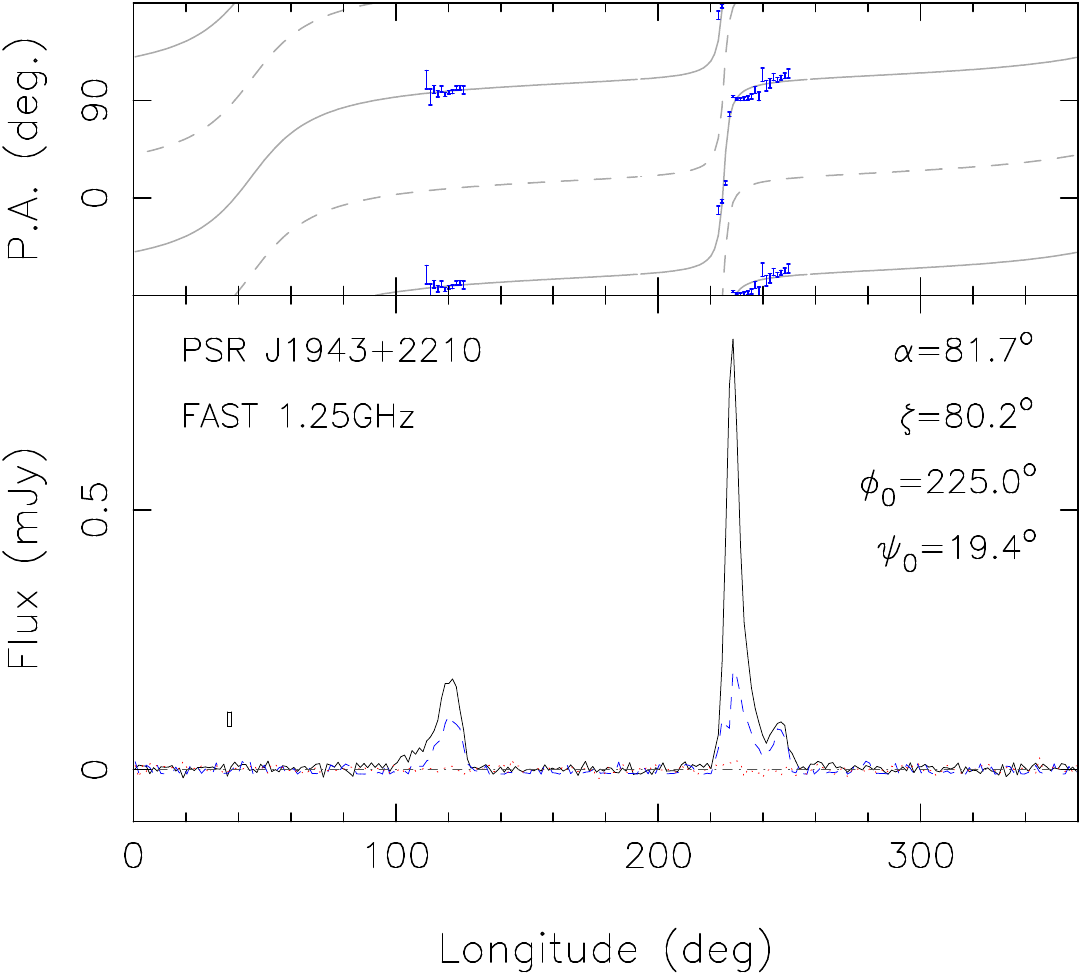}
    \includegraphics[width=0.32\textwidth]{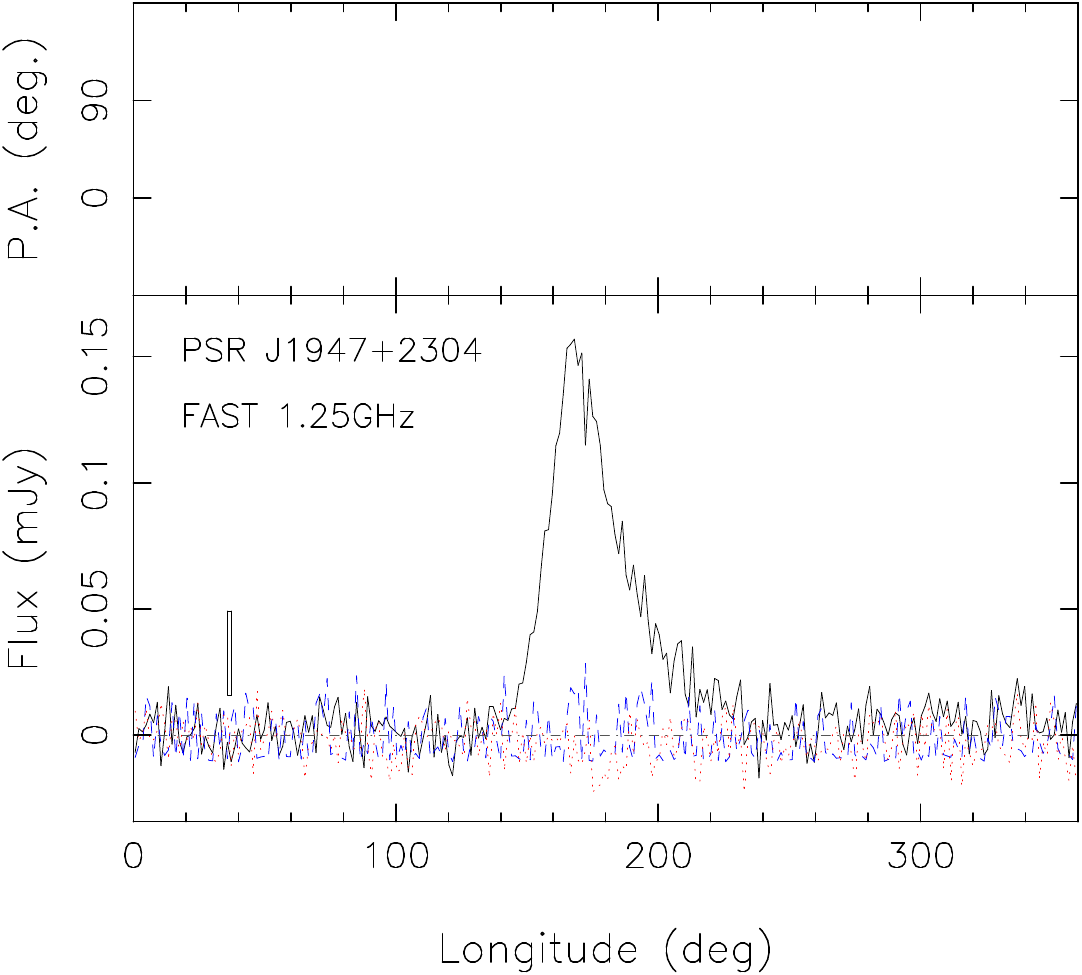}
    \includegraphics[width=0.32\textwidth]{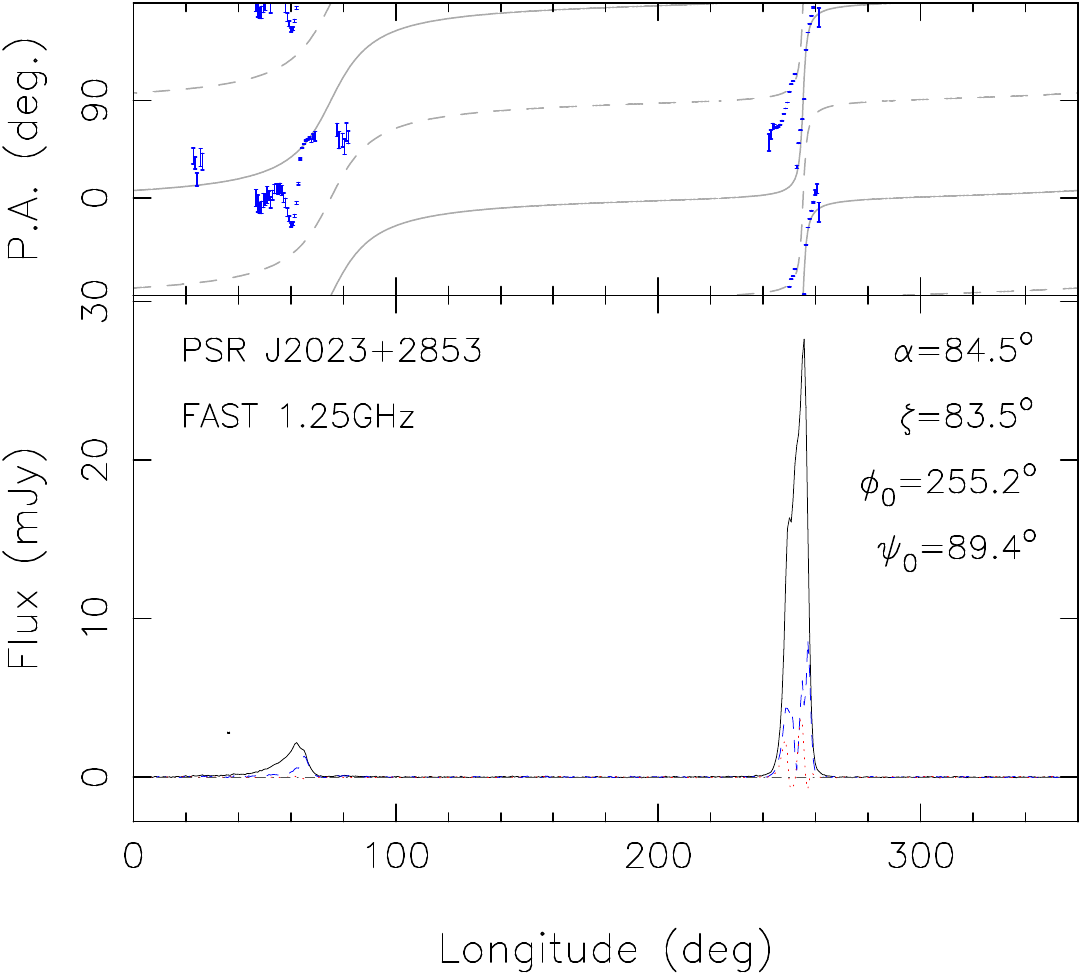}
    \caption{Polarization profiles of 6 pulsars. The total intensity profile (black line), linear (blue dash line), and circular (red dotted line, positive for the left-hand sense) polarization profiles are plotted in the lower sub-panel with a scale-mark for $\pm2\sigma$ and 1 bin-width, and the polarization angles (PA) are plotted in the upper sub-panel with the error bar for $\pm1\sigma$.  The best fitted PA curves by the rotating vector model \citep{Radhakrishnan+1969ApL.....3..225R} for PSRs J1943+2210, J0520+3722 and J2023+2853 are made with the ``psrmodel" command in the PSRCHIVE tool package \citep{Hotan+2004PASA...21..302H} and shown by the grey lines, together with the orthogonal mode by the grey dash lines. For PSRs J1919+1341 and J1947+2304, the rotation measure and polarized emission cannot yet be determined from available data.}
    \label{pol_profile}
\end{figure*}

\begin{figure*}
    \centering
    \includegraphics[width=0.24\textwidth]{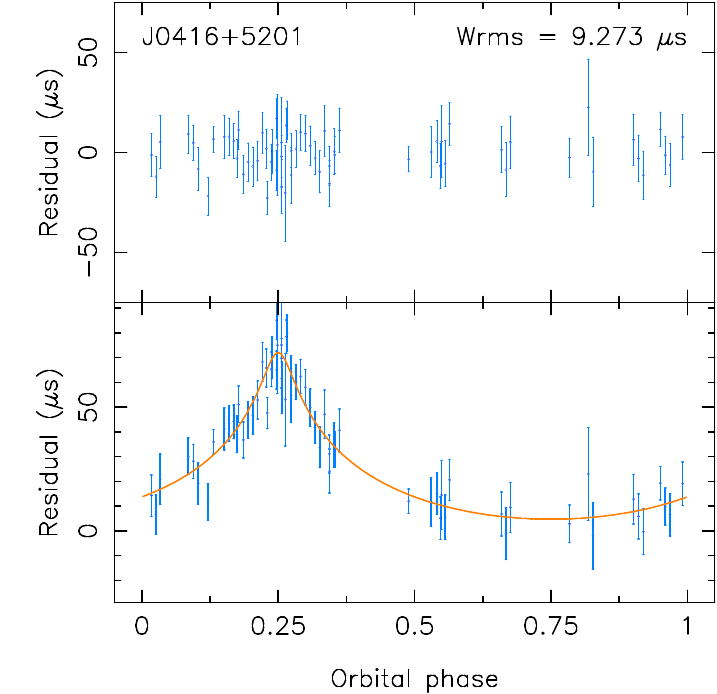}
    \includegraphics[width=0.24\textwidth]{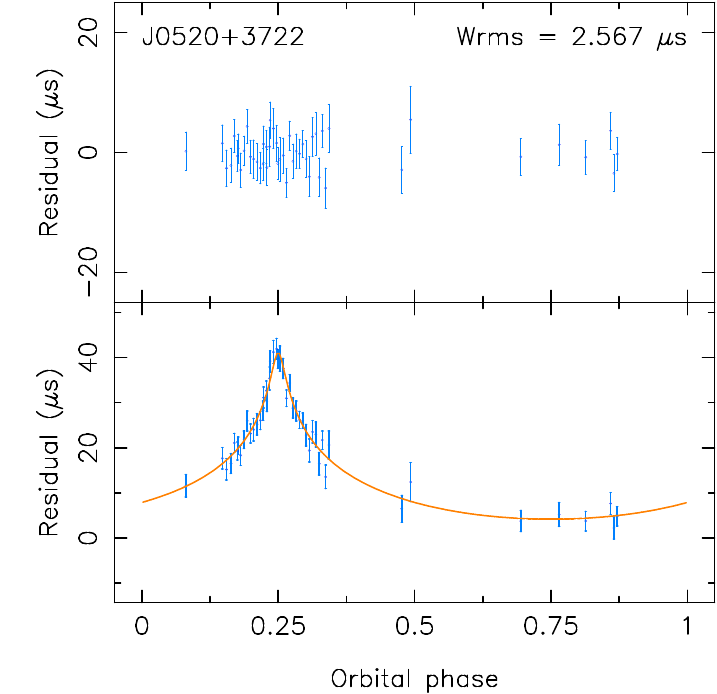}
    \includegraphics[width=0.24\textwidth]{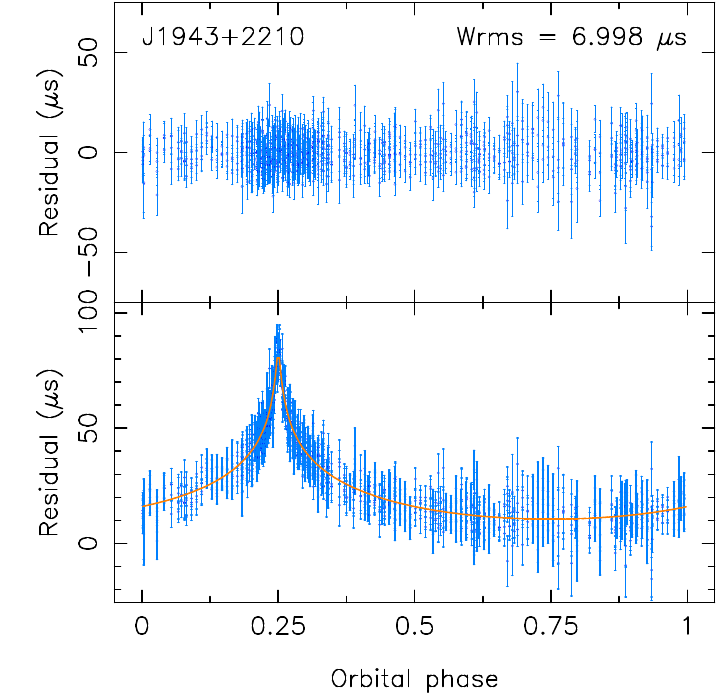}
    \includegraphics[width=0.24\textwidth]{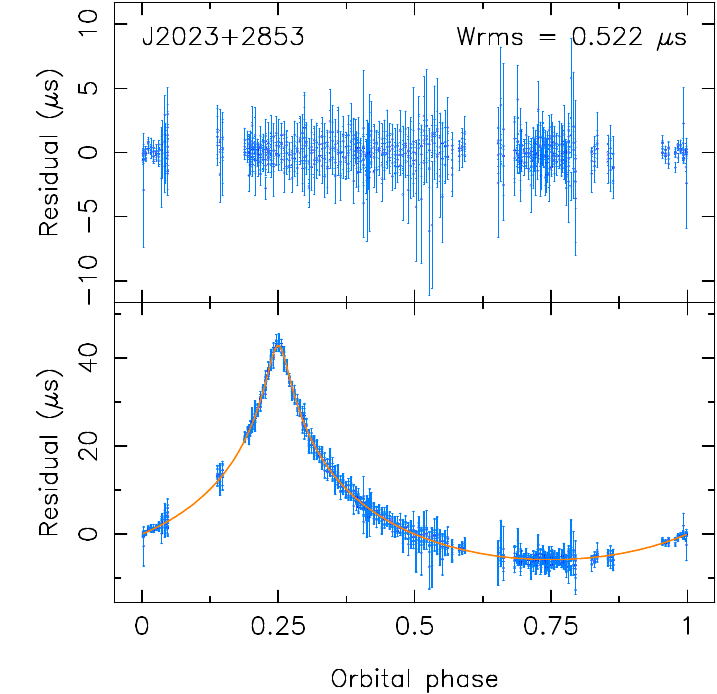}
    \caption{The detected Shapiro delay of PSRs J0416+5201, J0520+3722, J1943+2210 and J2023+2853 measured by FAST. In the upper sub-panels, the difference between measured TOAs and the derived timing model, $W_{\rm rms}$, are marked against orbital phase. In the lower sub-panels, we displayed the timing residuals after Shapiro delay is removed. All error bars represent for $\pm1\sigma$ uncertainty.}
    \label{timing_residuals-gpps0201}
\end{figure*}

\section{FAST Observations and Data Reduction}

All FAST observations have been carried out by using the L-band 19-beam receiver, covering a 500 MHz bandwidth centered at 1.25 GHz \citep{Jiang+2020RAA....20...64J}. The data were recorded for 2 or 4 polarization channels, with a sampling time of 49.152 $\mu$s for all 2048 or 4096 frequency channels. Before or after every observation session, the calibration noise signals with a modulated amplitude of 1.1~K and a period of 2.01326 s are injected into the receiver feeds so that polarization data can be calibrated \citep{Han+2021RAA....21..107H, wang+2023RAA}.

The six intermediate-mass binary pulsars, PSRs J0416+5201, J0520+3722, J1919+1341, J1943+2210, J1947+2304 and J2023+2853, 
were discovered during the FAST GPPS survey \citep{Han+2021RAA....21..107H, Han+2024RAA}. 
PSR J0416+5201 was discovered in a snapshot observation on 2023 January 3rd, and then confirmed in a 15-minute tracking observation on February 13th, 2023. Afterward, several follow-up observations are made and then the initial orbital parameters are determined. 
Similarly, PSR J0520+3722 was discovered from a snapshot observation on 2022 November 16th, and then confirmed in a 15-minute tracking observation on 2023 January 6th. 
PSR J1919+1341 was discovered from an observation on 2020 August 7th;
PSR J1943+2210 was discovered from an observation on 2021 June 27th;
PSR J1947+2304 was discovered from an observation on 2020 December 15th;
PSR J2023+2853 was discovered from an observation on 2019 March 29th \citep{Han+2021RAA....21..107H}.

For each pulsar, after a few observations have been done with the FAST,  we dispersed the data, and then searched for the barycenter spin periods and accelerations of the pulsar using PRESTO \citep{Ransom+2001PhDT.......123R}. These data are plotted in a two-dimensional plane and 
form an ellipse so that the orbital period ($P_{\rm orb}$) and the projected semi-major axis ($x$) can be obtained from the plot directly \citep{Freire+2001MNRAS.322..885F}. 

With this initial orbital period, we performed a two-dimensional search of orbital period ($P_{\rm orb}$) and time of ending node passage ($T_{\rm ASC}$) to fit the observed periods of different epochs, which provided the intrinsic spin period of the pulsar $P$ and refined the three orbital parameters $P_{\rm orb}$, $x$ and  $T_{\rm ASC}$ \citep{Bhattacharyya+2008MNRAS.387..273B}.

For timing analysis, firstly we folded pulsar profiles with the initial orbital parameters for data segments of 5 minutes each, using the package {\sc DSPSR} \citep{Straten+2011PASA...28....1V}. The ``times of pulse arrival'' (TOAs) were then extracted from these folded pulsar profiles using the package {\sc PSRCHIVE}, an Open Source C++ development library for the analysis of pulsar astronomical data \citep{Hotan+2004PASA...21..302H}. We then calibrate the polarization with the injected calibration signals using the command ``PAC'' in {\sc PSRCHIVE}, remove radio frequency interference using the ``2$\sigma$CRF'' method \citep{Chen+2023RAA....23j4004C}. The Faraday rotation measures (RMs) were then determined using ``RMFIT'' and then polarization data were RM-corrected using ``PAM''. The data of all frequency channels were then added to form profiles of four sub-bands. The profiles were compared to a frequency-average template produced by the command ``PSRSMOOTH'' or ``PAAS'', then finally TOAs were extracted from all these files by using the command ``PAT''.

\section{FAST Timing Results}

After FAST observations over years, including data obtained during the FAST GPPS survey and also by free-applied FAST projects (see details in authors contributions), and following the method in \citet{Freire+2018MNRAS.476.4794F}, we got the phase-connected timing solutions of these systems (see Table~\ref{timingsolution1} and \ref{timingsolution2}). Then, all data were refolded again to achieve better precision. If no dispersion measure (DM) variations are found, all frequency channels were summed together to obtain more precise TOAs. The timing residuals are presented in Fig.~\ref{timing_residuals_other} and Fig.~\ref{timing_residuals-gpps0201}. No systematic trends as a function of epoch or orbital phase can be seen in these plots. During the analyses, the Solar System ephemeris DE436 \footnote{https://naif.jpl.nasa.gov/pub/naif/JUNO/kernels/spk/de436s.bsp.lbl} was applied to correct the motion of the FAST relative to the barycenter of the Solar System, while the motion of the pulsars is described using the ELL1 model (or ELL1k model for Shapiro delay parameters) \citep{Lange+2001MNRAS.326..274L, Edwards+2006MNRAS.372.1549E}, which are the commonly-used orbital model for low eccentricity orbits. 

We also divided the FAST observation data of the 500 MHz bandwidth into 4 sub-bands and then refine DM and RM values. After final corrections, we get the final polarization profiles as shown in Fig.~\ref{pol_profile}. Linear polarizations of pulse profiles of PSRs J1919+1341 and J1947+2304 are not yet detected from available data. Discussions of polarization profiles and geometry are presented in Sect.4.3. The distances of these pulsars are estimated from the DMs by using the electron-density Model NE2001 \citep{ne2001} and YMW16 \citep{ymw16}. Based on the precise astronomic positions of these pulsars from timing observations, we searched for their optical counterparts in the Pan-STARRS1 image but yielded null result \citep{Chambers+2016arXiv161205560C}.

Among the six pulsars, we get the Shapiro delay for PSRs J0416+5201, J0520+3722, J1943+2210 and J2023+2853, which indicates that they are in edge-on orbits. When the orbit of a pulsar is nearly circular, the Shapiro delay can be described by PK parameters $r\equiv Gm_{\rm c}/c^3$ and $s\equiv\sin i$ via  
\begin{equation}
    \Delta_{\rm s}=-2r\ln\{1-s\cos[2\pi(\phi-\phi_{\rm o})]\},
\end{equation}
here $m_{\rm c}$ is the companion mass, $i$ is the orbital inclination angle, $\phi$ is the orbital phase and $\phi_{\rm o}$ is the conjunction phase at $\phi=0.25$. Together with the mass function, 
\begin{equation}
f(m)=(4\pi^2/G)x^3/P_{\rm orb}^2=(m_{\rm c}\sin i)^3/(m_{\rm p}+m_{\rm c})^2,
\end{equation}
the observed Shapiro delay can constrain the pulsar mass $m_{\rm p}$, the companion mass $m_{\rm c}$, and the orbital inclination $i$.

In the following, we discuss the detailed timing results of 6 pulsars.

\subsection{PSR J0416+5201}

PSR J0416+5201 has a spin period of 18.2 ms. The pulse profile consists of the one-component main pulse and double-peak interpulse with a separation of about 180$^\circ$ in the rotation longitude (see Figure~\ref{pol_profile}). 

Assuming the pulsar mass of 1.4 $M_\odot$, the minimum companion mass,  derived from mass function, should be 1.28 $M_\odot$, in the mass range of neutron stars. But such a neutron star companion is rebutted by the circular orbit with an eccentricity of only a few 10$^{-5}$.  Only an ONeMg WD can have such a mass. On the other hand, a WD should have a mass less than the Chandrasekhar mass limit of $\sim$1.4 $M_\odot$. We therefore can conclude that the companion mass of PSR J0416+5201 is probably in the range of 1.28 and 1.4 $M_\odot$. Previously discovered pulsars with such massive WD companions are PSRs J1227$-$6208 \citep[$1.21<m_{\rm c}$/M$_\odot<1.47$; see][]{Colom+2024A&A...690A.253C}, J2222$-$0137 \citep[m$_{\rm c}=1.319(4)$ $M_\odot$; see][]{Guo+2021A&A...654A..16G}, J1528$-$3146 \citep[m$_{\rm c}=1.33_{\,‑0.07}^{+0.08}$ $M_\odot$; see][]{Berthereau+2023A&A...674A..71B}, J1439$-$5501 \citep[m$_{\rm c}=1.27_{\,‑0.12}^{+0.13}$ $M_\odot$; see][]{Jang+2024A&A...689A.296J} and B2303+46 \citep[$1.2<m_{\rm c}$/M$_\odot<1.4$; see][]{van+1999ApJ...516L..25V}, but non of which have a compact orbit like PSR J0416+5201. The massive companion in an orbital period of 0.396 days should merge with PSR J0416+5201 in about 3.1 Gyr. 

We detected the Shapiro delay of the PSR J0416+5201 binary system, which implies an edge-on orbit. The companion mass is then further constrained by the Shapiro delay as being 1.7(5) $M_\odot$ (see Table~\ref{timingsolution1}), which provides an independent coarse but consistent mass constraint.

\subsection{PSR J0520+3722}

The phase-connected timing solution of this binary system has been obtained from the timing data over two years.
Among six IMBPs, PSR J0520+3722 has the fastest spin period of 7.91 ms and the youngest characteristic age of 0.35 Gyr. Its pulse profile has a main pulse and an interpulse with a separation of about 180$^\circ$ in the rotation longitude (see Figure~\ref{pol_profile}), with a clear circular polarization sense-reversal at the center of the two-peak main pulse. 

The companion mass is constrained by the mass function of the binary and the Shapiro delay detected in timing data, as being 0.70(13) $M_\odot$, indicating that its companion is a CO WD. However, the precision of two measurable Shapiro delay parameters is not precise enough to constrain the pulsar mass. Assuming a pulsar mass of 1.4 $M_\odot$, the pulsar will merge with its companion after $\sim12$ Gyr.

\subsection{PSR J1919+1341}

PSR J1919+1341 is a mildly recycled pulsar with a spin period of 11.7 ms. 
PSR J1919+1341 has been frequently detected by the FAST L-band 19-beam receiver even when the FAST is tracking PSRs J1918+1340g (GPPS0081) and PSR J1920+1340g (GPPS0423), though it is 1.5~arcmin offset from the beam-center of one of the 19 beams. The TOAs have a large RMS timing residual of 85 $\mu$s due to the weaker pulse, the beam-offset, the wide pulse and also the relatively longer spin period. 
Assuming the pulsar mass being 1.4 $M_\odot$, the minimum mass of its companion is 0.78 $M_\odot$. Because of the very circular orbit, PSR J1919+1341 is probably circling a massive WD companion with an orbit period of 0.370 days. The merger timescale of this system is estimated to be 4.7 Gyr. 
Large timing residuals prevent this system from having a good measurement of the Shapiro delay. Without any further constraints on the companion mass, its companion can be either a CO WD or an ONeMg WD.

\begin{figure*} 
    \centering
    \includegraphics[width=0.7\columnwidth]{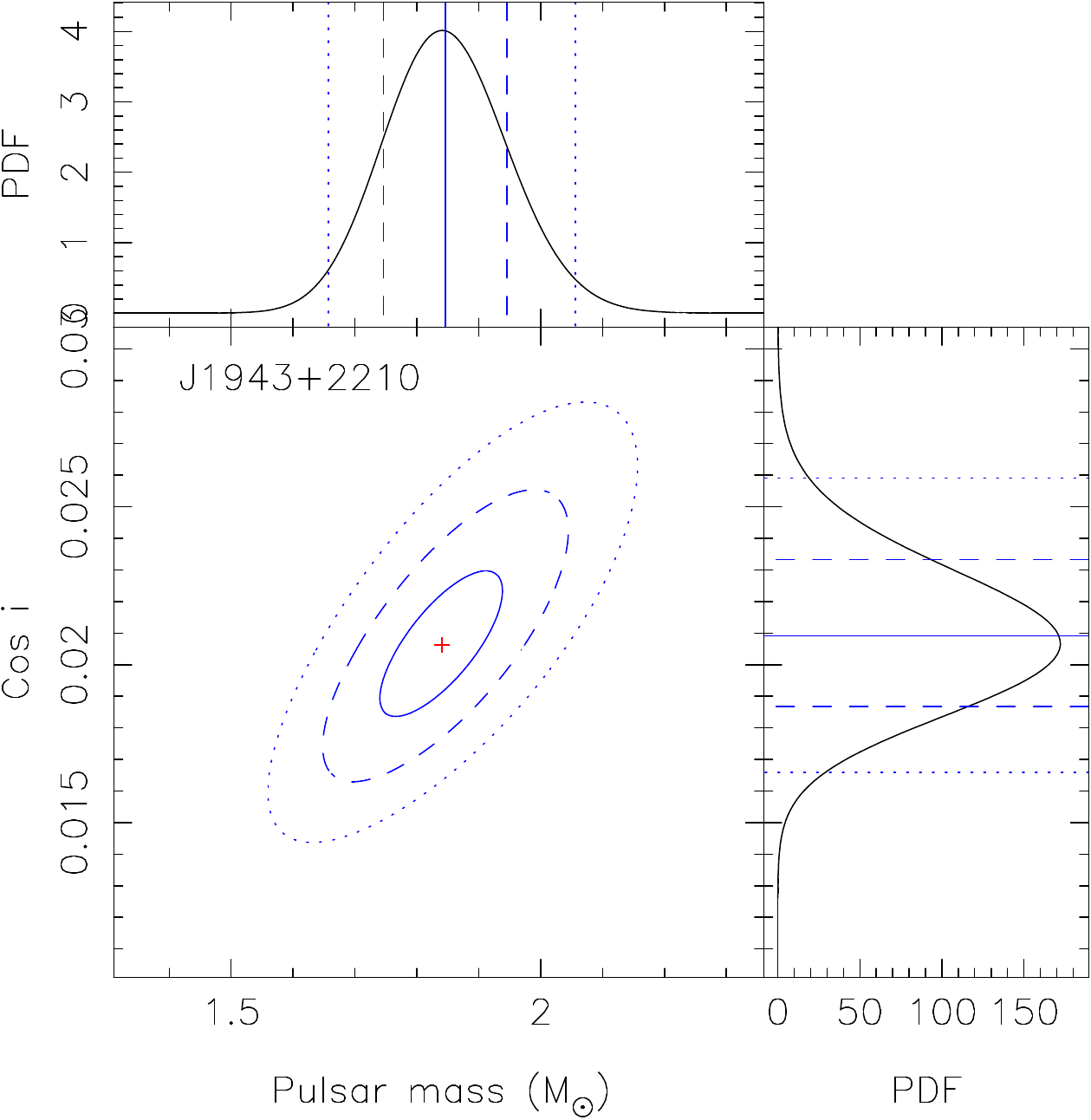} \hspace{10mm}
    \includegraphics[width=0.7\columnwidth]{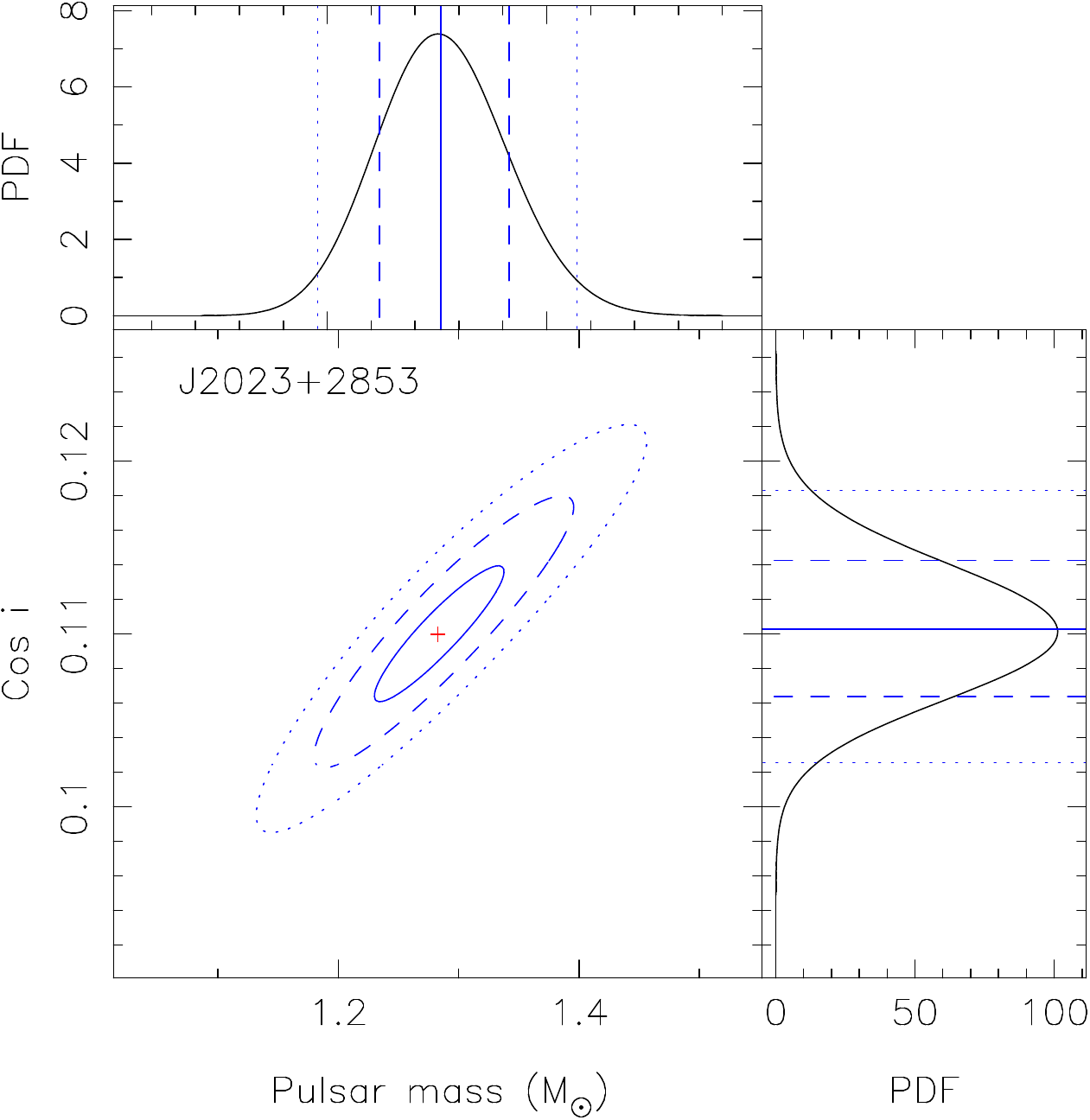} 
    \caption{Constraints on the pulsar mass $m_{\rm p}$ and orbital inclination angle $i$ for PSRs J1943+2210 and J2023+2853 by using the measured Shapiro delay. The contours in the main sub-panels show the likelihood in the 2-dimensional probability distribution function (PDF) at the 39\% ($\Delta\chi^2=1$), 86\% ($\Delta\chi^2=4$), and 99\% ($\Delta\chi^2=9$) confidence level for the fitted ELL1k binary model. The red crosses indicate the best masses for the minimum $\chi^2$. The 1-dimensional PDFs for the $m_{\rm c}$ and $\cos$i are shown in the upper and right sub-panels with indications for the median, 1$\sigma$, and 2$\sigma$ ranges.}
    \label{Bayesian analysis}
\end{figure*}

\subsection{PSR J1943+2210}

After 3 years of timing observations, we get the phase-coherent timing solution. PSR J1943+2210 has a spin period of 12.9 ms in a compact orbit with a period of 0.372 days and an eccentricity of 2.44(95)$\times$10$^{-6}$. The companion has a minimum mass of 0.89 $M_\odot$ according to the mass function. Based all these measurements, we suggest that the companion should be a massive WD. The pulsar has a double-peak main pulse and a discrete one-peak secondary pulse with a rotation longitude separation of about 100$^\circ$, so it is not the interpulse (see Figure~\ref{pol_profile}).

 We detected the proper motion of 7.4(0.5) mas\,yr$^{-1}$. Considering the DM distance of 4.6 kpc \citep{ne2001} or 4.0 kpc \citep{ymw16}, one can get the transverse velocity of $\sim1.5\times10^{2}$ km\,s$^{-1}$, which is high 
but still reasonable \citep{Shamohammadi+2024}.  Combining the mass function of this binary system and the Shapiro delay (see Figure~\ref{Bayesian analysis}), we get the companion mass as being 1.03$^{\,+0.04}_{-0.03}$ and an orbital inclination being 88.80$^{\,+0.13}_{-0.14}$. The so-derived NS mass is 1.84$^{\,+0.11}_{-0.09}$ $M_\odot$, one of the largest birth masses ever known \citep{Tauris+2023pbse.book.....T}.The measured companion mass is in the conjunction area between the CO WDs and ONeMg WDs, so it can be either a CO WD or an ONeMg WD.

Because of a relatively short orbital period and a high NS mass, this binary system has a coalescence time of 2.5 Gyr, the shortest among those IMBP systems with a recycled pulsar.

\subsection{PSR J1947+2304}

Using FAST observations of PSR J1947+2304 over more than 3 years, we get the phase-connected timing solution of this binary pulsar. It has a spin period of 10.9 ms, and an orbit period of 0.339 days, with an eccentricity of a few 10$^{-5}$. The companion has a minimum companion mass of 0.87 $M_\odot$. We cannot get other constrains on the companion mass, and hence cannot determine the companion type as being a CO WD or an ONeMg WD.

Among all binary MSPs with a massive WD companion, the orbital period of PSR J1947+2304 is the shortest. Such a tight orbit could be the result of continuous gravitational wave emission over a long time if expressible by the pulsar characteristic age of 3.6 Gyr. General relativity predicts a variation rate of orbital period $\Dot{P}_{\rm orb}\sim-1\times10^{-13}$, which should be measurable in the future. This binary system will merge within 3 Gyr due to gravitational wave emission.

\subsection{PSR J2023+2853} 

PSR J2023+2853 is a very bright pulsar discovered in the FAST GPPS survey, which has been buried in the harmonics of the nearby extremely bright pulsar PSR B2020+28 with a similar DM \citep{Han+2021RAA....21..107H}. It has a spin period of 11.3 ms and is in a binary system with an orbital period of 0.72 days. Its pulse profile has an interpulse (see Figure~\ref{pol_profile}), separated from the main pulse by 180 degrees in the rotation longitude. 

The orbit of PSR J2023+2853 is highly inclined, so the Shapiro delay can be very well detected and described using the PK parameters of $r\equiv Gm_{\rm c}/c^3$ and $s\equiv\sin i$. On January 17th, 2024 we launched a tracking observation for 3.5 hours by the FAST to cover the conjunction phase and a clear Shapiro delay was detected. 
When we drafted this paper, we noticed that the CHIME team published the timing results on the arXiv on 2024 February 13 for some bright pulsars with high cadences, including this bright pulsar \citep{Tan+2024ApJ...966...26T}. 
They obtained the two Shapiro delay parameters and obtained a companion mass of 0.93$^{\,+0.17}_{-0.14}$ $M_\odot$ and a pulsar mass of 1.50$^{\,+0.49}_{-0.38}$ $M_\odot$ \citep{Tan+2024ApJ...966...26T}. 

The timing solutions we obtained by FAST are consistent with their published results but with much better precision. From the two PK parameters $r$ and $s$ we obtained $m_{\rm c}$=0.85(2) $M_\odot$ and $\sin i$=0.9939(4). 
Because of the long span for timing and the better precision of measurements, we have also measured the proper motion of this pulsar as being 8.745(77) mas\,yr$^{-1}$. The parallax is 1.22(30) mas, which should be counted when used for a test of general relativity. The corresponding parallax distance is 0.8(0.2) kpc, closer than but still consistent with its DM distance estimate within 3$\sigma$ uncertainty.
All these measurements can constrain the masses of PSR J2023+2853 and its companion. We performed the Bayesian analysis of parameters, as displayed in Fig.~\ref{Bayesian analysis}. We get the pulsar mass of $m_{\rm p}$ of 1.28$^{\,+0.06}_{-0.05}$ $M_\odot$, the companion mass of $m_{\rm c}$ of 0.853$\pm0.020$ $M_\odot$, the total mass $m_{\rm tot}$ of 2.14$^{\,+0.08}_{-0.07}$ $M_\odot$, and an orbital inclination $i$ of 83.7$^{\,+0.2}_{-0.3}$ deg. Similar results have been obtained using the DDGR model-fitting \citep{Damour+1985AIHPA..43..107D, Damour+1986AIHPA..44..263D}, which gives the companion mass of 0.857$\pm$0.018 $M_\odot$ and the total mass of 2.15$\pm$0.07 $M_\odot$. According to the companion mass, its companion is a CO WD.

\section{Discussions}

These 6 compact IMBP systems should be products of CE evolution \citep{Tauris+2011MNRAS.416.2130T, Tauris+2012MNRAS.425.1601T, Tauris+2023pbse.book.....T}. The mass accretion should cause the spin axis aligned with the orbit axis. These two compact objects with different masses can act as a laboratory to test gravitational theories. The precise timing of such systems can measure the pulsar masses which are the key parameters for understanding the evolution channels of neutron stars and are constraints for  the equation of state of supra-nuclear matter \citep[e.g.][]{Horvath+2021AN....342..294H}.

\begin{figure}[htp]
    \centering
    \includegraphics[width=0.95\columnwidth]{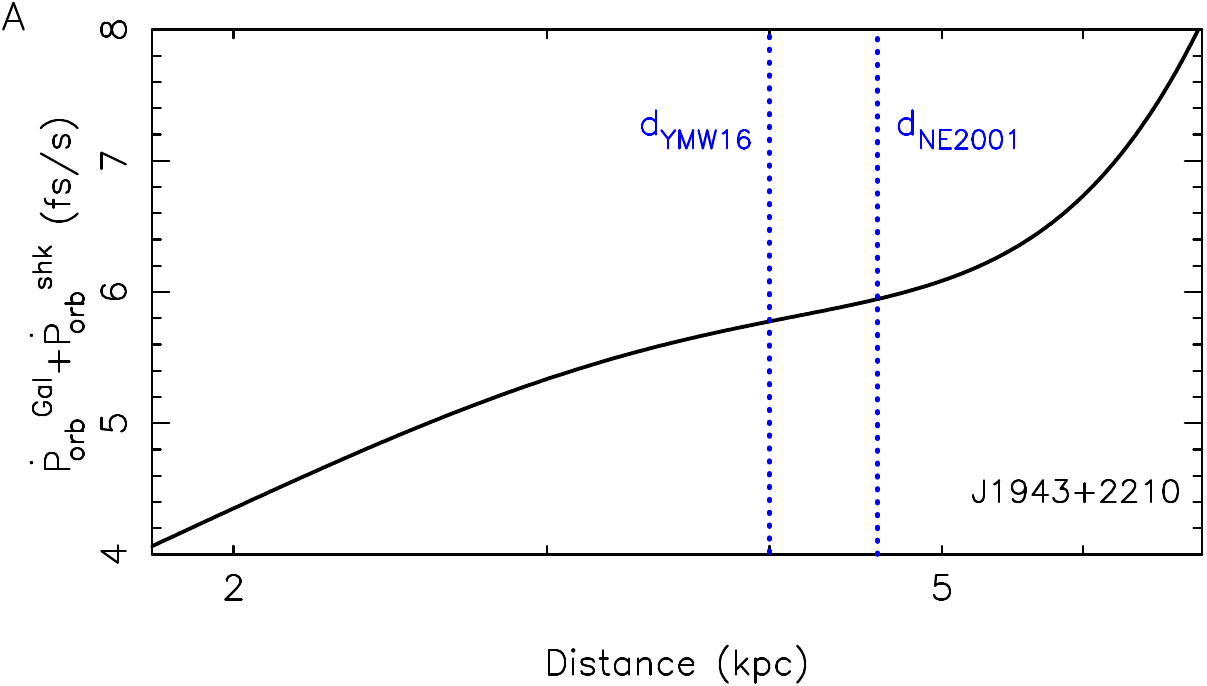}
    \includegraphics[width=0.95\columnwidth]{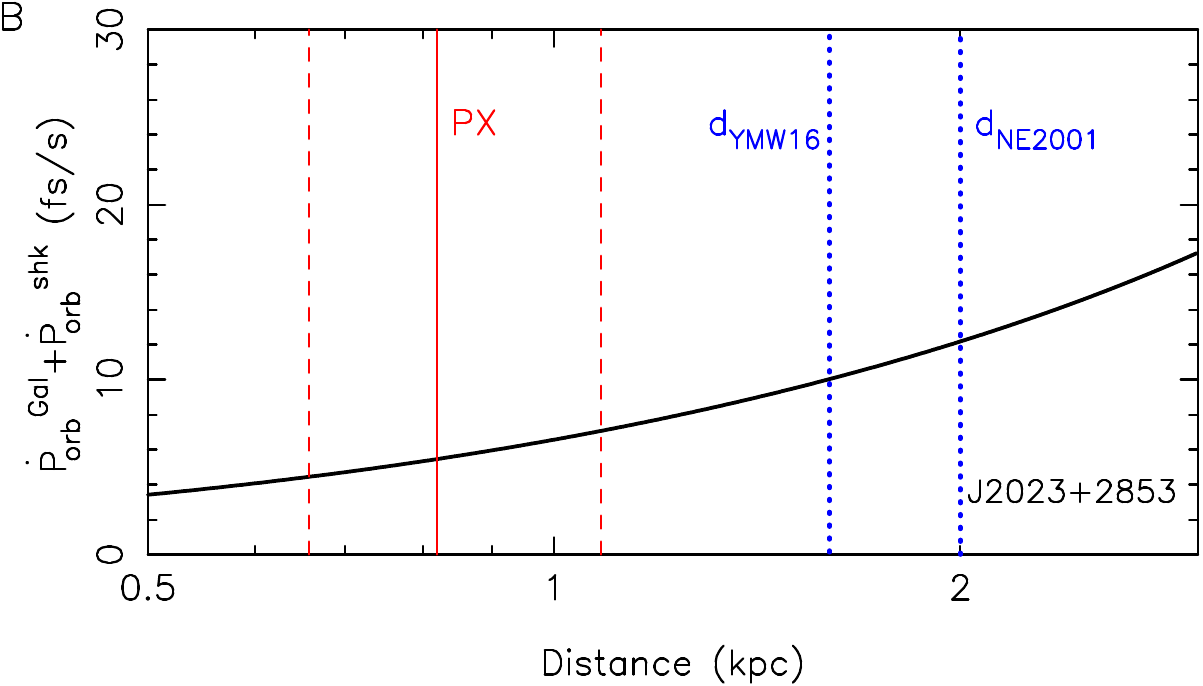}
    \caption{The sum of the predicated observed orbital decay due to the Galactic acceleration and the proper motion as a function of distance to the pulsar. The black line shows the variation of kinematic contribution to the $\dot{P}_{\rm orb}$ as a function of distance. The blue dotted lines indicate the estimated distances by using the Galactic electron density model of the NE2001 \citep{ne2001} and YMW16 \citep{ymw16}. The pulsar distance of PSR J2023+2853 has been derived from the timing parallax, with the 1 $\sigma$ uncertainty indicated by the vertical dashed lines.}
    \label{PBdot excess}
\end{figure}

\subsection{Possible tests of the gravitational theory with high accuracy data of PSRs J1943+2210 and J2023+2853}

FAST measurements of PSRs J1943+2210 and J2023+2853 give an rms residual with a high precision of 7 and 0.5 $\mu$s respectively. Because of highly inclined orbits, the two Shapiro delay parameters have been used to constrain the companion masses, pulsar masses, and the inclination angles of the orbits. Their precise proper motions and the parallax of J2023+2853 are also measured from timing data, as shown above. The periastron advance ($\Dot{\omega}$) and changes of orbital period $\overset{.}{P}_{\rm orb}$ are still not well measured, but their uncertainties will decrease with a longer observation span $T$ by $T^{-3/2}$ and $T^{-5/2}$, respectively. 

In some gravity theories, such as the Scalar-tensor theories, the dipolar gravitational waves are predicted to radiate from an asymmetric binary system with significantly different masses with different gravitational self-energies, though not in general relativity. Binary millisecond pulsars with a white dwarf companion indeed provide a unique opportunity for such a test \citep{Freire+2012MNRAS.423.3328F, Guo+2021A&A...654A..16G, Gautam+2022A&A...668A.187G}. The orbital period derivatives $\dot{P}_{\rm orb}$ of PSRs J1943+2210 and J2023+2853 will be especially useful to constrain the spontaneous scalarization \citep{Damour+1993PhRvL..70.2220D}, as shown for the case of PSR J2222-0137 by \citet{Zhao+2022CQGra..39kLT01Z}. According to \citet{Cognard+2017ApJ...844..128C}, for the binary orbit with a negligible eccentricity, the additional change rate of the orbital period in the Scalar-tensor theories should be 
\begin{equation}
    |\alpha_{\rm p}-\alpha_{\rm c}|^2=\delta\Dot{P}^{\rm\,D}_{\rm orb}(\frac{P_{\rm orb}}{4\pi^2})(\frac{M_\odot}{T_\odot m_{\rm c}})(\frac{q+1}{q}),
\end{equation}
where $\alpha_{\rm p}$ and $\alpha_{\rm c}$ is the scalar charge or effective coupling strength of the pulsar and its companion, $\delta\Dot{P}^{\rm\,D}_{\rm orb}$ is the orbital damping caused by the dipolar gravitational wave emission, and $q$ is the mass ratio $m_{\rm p}/m_{\rm c}$. Obviously the theoretical constrains for $|\alpha_{\rm p}-\alpha_{\rm c}|^2$ depends on the precision of $\delta\Dot{P}_{\rm orb}^{\rm D}$. For PSR J1943+2210, if $\delta\Dot{P}_{\rm orb}^{\rm D}$ can be constrained to be less than 4\,fs/s, then $|\alpha_{\rm p}-\alpha_{\rm c}|$ can reach $1\times10^{-3}$. For PSR J2023+2853, if we can constrain $\delta\Dot{P}_{\rm orb}^{\rm D}$ to be less than 1\,fs/s, then $|\alpha_{\rm p}-\alpha_{\rm c}|$ should be smaller than $1\times10^{-3}$, better than the constrains from other systems, e.g.  $1.9\times10^{-3}$ for PSR J1738+0333 from \citet{Freire+2012MNRAS.423.3328F} and $3.3\times10^{-3}$ for PSR J2222$-$0137 from \citet{Guo+2021A&A...654A..16G}. We do need to observe it for a longer observation span $T$.

However, the observed excessive orbital period derivative may have other possible origins, as expressed by  
\begin{equation}
    \Dot{P}^{\rm\,ex}_{\rm orb}=\delta\Dot{P}^{\rm\,D}_{\rm orb}+\Dot{P}^{\rm\,Gal}_{\rm orb}+\Dot{P}^{\rm\,shk}_{\rm orb},
\end{equation}
here $\Dot{P}^{\rm\,Gal}_{\rm orb}$ is the contribution of Galactic acceleration, $\Dot{P}^{\rm\,shk}_{\rm orb}$ is caused by the proper motion \cite{sh70}. Taking the distance from the Sun to the Galactic center as being 8.0~kpc and the Galactic circular velocity of 220 km~s$^{-1}$ at the location of the Sun  \citep{GRAVITY+2019A&A...625L..10G}, we find that the sum of the two contributions is a function of distance, as shown in Fig.~\ref{PBdot excess}. 

At present, no reliable distance measurement for PSR J1943+2210 prevents the accuracy of kinematic effects and hence a  good constraint on the dipolar gravitational wave emission (see Figure~\ref{PBdot excess}). Based on the parallax distance of PSR J2023+2853, the sum of the two contributions from the Galactic acceleration and the proper motion should be 4.4--7.1\,fs/s in the 68\% confidence level. In the future, with more data for more accurate measuring parallax (for PSR J2023+2853) and the Shapiro delay and proper motion, the two binary systems of PSRs J1943+2210 and J2023+2853 can potentially provide the best constraint on the existence of dipolar gravitational wave emission.

\subsection{Two possible formation channels of compact IMBP systems}

If the white dwarf companion was formed before the neutron star, the binary system should be in an eccentric orbit \citep{Antoniadis+2011MNRAS.412..580A}, as indicated by the filled triangle in Fig.~\ref{pb-m2}. Otherwise, the orbits of the binary systems should be nearly circular \citep{Tauris+2023pbse.book.....T}. 

All six IMBPs presented in this paper have a circular orbit, therefore these systems probably are descents from IMXBs with a very wide orbit of $P_{\rm orb}\sim10^2$--$10^3$~days. The companion star ever had a mass of 2 -- 10 $M_\odot$, and it became a giant star. Such a giant star responds to the mass loss by expanding, and then the mass transfer rate arises quickly even far exceeding the mass accretion rate of the NS, then a CE is formed. The NS will spiral into the CE during which orbital momentum, and the dynamic energy of the NS is then transferred into the CE. If the CE is successfully ejected in $\lesssim10^3$ yr \citep{Podsiadlowski+2001ASPC..229..239P, Passy+2012ApJ...744...52P, Tauris+2012MNRAS.425.1601T}, the outcome of this spiral-in phase should be a compact binary system consisting of a NS and a He star \citep{Paczynski+1976IAUS...73...75P, Iben+1993PASP..105.1373I, Ivanova+2013A&ARv..21...59I}. 

The accreted mass during the CE phase can recycle the neutron star to a spin period of $\sim10$ ms, with an amount of about 0.01 $M_\odot$. Conventionally, such spin periods should be reached in the post-CE mass accretion phase triggered by the Case BB RLO, during which the mass is transferred from a giant helium star to an NS, but it takes a much longer time of $\sim10^5$ yr to get partially recycled \citep{Tauris+2011MNRAS.416.2130T, Tauris+2012MNRAS.425.1601T}. Such 
a Case BB RLO evolution phase can only happen for those NS plus He star binaries with a tight orbit, in which the orbital energy of the progenitor is sufficient to fully eject the envelope of the donor star that evolves to near the tip of the red giant branch. A compact pulsar-helium star binary system should be left after the CE ejection. To avoid a merger, the orbital period should be $\lesssim$ 0.5 days\citep{Tauris+2011MNRAS.416.2130T, Tauris+2012MNRAS.425.1601T}. After the Case BB RLO phase, a compact IMBP system with a mildly recycled pulsar should be formed. 

Another possible formation channel for these compact IMBPs is the Case C channel, where the CE phase was initiated with an asymptotic giant branch star. Unlike the Case BB evolution channel, the helium core of the companion star has been exhausted, and the system will not go though Case BB RLO. The hypercritical accretion cooled by neutrino loss during CE phase \citep{Houck+1991ApJ...376..234H} or post-CE thermal remnant thermal readjustment phase \citep{Ivanova+2011ApJ...730...76I} may significantly spin up the NS, providing an alternative explanation to the mildly recycled pulsars. To explain the heavy NS mass 1.84$^{\,+0.11}_{-0.09}$ $M_\odot$ of PSR J1943+2210 that has a spin period of 12.9 ms, the accreted mass is about $\sim 0.01~M_\odot$ according to equation (14) in \cite{Tauris+2012MNRAS.425.1601T} in the recycled process, so the initial mass of PSR J1943+2210 should be about 1.83 $M_\odot$.  It is a born-massive NS,  similar to those in the systems of PSRs J1614$-$2230, PSR J1640+2224, PSR J2222$-$0137 and 2A 1822$-$371 \citep{Tauris+2011MNRAS.416.2130T, Deng+2020ApJ...892....4D, Guo+2021A&A...654A..16G, Wei+2023A&A...679A..74W}.

Such compact IMBP systems are rare. Previously only five MSPs with a massive WD companion are in compact orbits with a period less than 1 day \citep{J1757-5322+Edwards2001, J1802-2124+Faulkner2004, J1748-2446N+Ransom2005, Manchester+2005AJ....129.1993M, J1952+2630+Knispel2011, J1525-5545+Ng2014}, as listed in Table~\ref{compact IMBP}. We find six compact IMBPs with orbital periods less than 1 day, PSRs J0416+5201, J0520+3722, J1919+1341, J1943+2210, J1947+2304 and J2023+2853. Five of them will merge within a Hubble time except for the last one, making them the progenitors of the Galactic gravitational wave sources for space-borne gravitational wave detectors like Tianqin \citep{Luo+2016CQGra..33c5010L}, Taiji \citep{Ruan+2020IJMPA..3550075R}, and LISA \citep{Amaro-Seoane+2017arXiv170200786A}. 
Their tight orbits, relatively fast spin periods and the large companion masses are all consist with the prediction of the Case BB RLO evolution channel \citep{Tauris+2011MNRAS.416.2130T, Tauris+2012MNRAS.425.1601T, Tauris+2023pbse.book.....T}.

\subsection{Geometry and pulse profile}

During the Case BB RLO, the NS is spun up so that the rotation axis should get aligned with the orbital angular momentum \citep{Yang_2023AAS}. If so, and if the orbit is nearly edge-on, then our line of sight should be perpendicular to the pulsar spin axes. Noticed that the Shapiro delay signals have been detected in PSRs J0416+5201, J0520+3722, J1943+2210 and J2023+2853, which is a clear indication for their edge-on orbit. The detected radio emission indicates that their magnetic field poles are aligned to our line of sight. If the surface magnetic fields of such a pulsar are is dipolar, then the magnetic axis must be highly inclined from the spin axis. So they are perpendicular rotators, from which an interpulse may be observed.  Indeed, all these pulsars PSRs J0416+5201, J0520+3722, J1943+2210 and J2023+2853 have a pulse profile with two discrete pulses. Among them an interpulse is observed approximately half-way the spin cycle except for PSR J1943+2210. 

PSRs J1913+1341 and J1947+2304 only have one pulse and no Shapiro delay signals have been detected yet. Their orbital inclinations are not constrained yet due to large TOA uncertainties.

Though the rotating vector model \citep{Radhakrishnan+1969ApL.....3..225R} has been widely applied to normal pulsars, we tried to fit the model to PSRs J0520+3722, J1943+2210 and J2023+2853 and get large inclinations of magnetic axis from the rotation axis, see $\alpha$ and $\zeta$ in Fig.~\ref{pol_profile}. These results suggest that the neutron stars, i.e. the pulsar, have been spun up by stable mass transfer process in e.g. the Case BB RLO \citep{Tauris+2023pbse.book.....T} or post-CE thermal remnant thermal readjustment phase \citep{Ivanova+2011ApJ...730...76I}. If the system was formed via the Case C channel and the NS was spun up via the neutrino- cooling accretion during the CE phase, the pulsar spin axis may be misaligned with the orbital orientation. 

Among other known compact IMBPs, PSRs J1525$-$5545, J1757$-$5322 and J1748$-$2446N have only one on-pulse region and no constraint on their orbital inclination angles has been reported yet \citep{J1757-5322+Edwards2001,J1525-5545+Ng2014,J1748-2446N+Ransom2005}. PSRs J1802$-$2124 and J1952+2630 also have only one on-pulse region, and the orbital inclination of J1802$-$2124 is 79.9(6) deg and that of J1952+2630 is $\sim$72 deg \citep{Ferdman+2010ApJ...711..764F, Gautam+2022A&A...668A.187G}. More detail study of their geometry may help us to understand their origin.

\section*{Acknowledgements}
We thank Dr Miquel Colom i Bernadich, Dr Paulo Freire and the two referees for their careful reading of the manuscript and helpful suggestions. This work made use of the data from FAST (Five-hundred-meter Aperture Spherical radio Telescope) (https://cstr.cn/31116.02.FAST). FAST is a Chinese national mega-science facility, operated by National Astronomical Observatories, Chinese Academy of Sciences. The GPPS survey project is one of five key projects of FAST. The authors are supported by the National Natural Science Foundation of China (NSFC, Grant Nos. 11988101, 12133004 and 11833009) and the Research Program of the Chinese Academy of Sciences (grant No. QYZDJ-SSW-SLH021 and JZHKYPT-2021-06).

\section*{Authors contributions}  
The FAST GPPS survey is a key FAST science project led by J.~L. Han. He organized the teamwork for the survey and follow-up observations,  processed the survey data, and discovered these pulsars. Timing data have been observed by P. F. Wang (via PT2023\_0190), T. Wang (via PT2022\_0159),  W. Q. Su (PT2021\_0132), Chen Wang (via PT2023\_0017),  Z. L. Yang (via PT2023\_0084 and PT2024\_0200) and D. J. Zhou (via PT2023\_0188). 
Z.~L. Yang processed all data presented in this paper and drafted the manuscript under the supervision of J.~L. Han. 
J.~L. Han was finally in charge of finishing this paper.
P.~F. Wang developed the processing procedures for pulsar polarization profile and pulsar timing which are extensively used in this paper.
Chen Wang feeds all targets for the GPPS observations. 
D.~J. Zhou, Tao Wang, W.~C. Jing, Yi Yan, Lang Xie and  N.~N. Cai contributed to different aspects of data processing and/or joined many group discussions.
P.~F. Wang and Jun Xu made fundamental contributions to the construction and maintenance of the computation platform.
Other people jointly propose or contribute to the FAST key project.
All authors contributed to the finalization of this paper.  \\


\section*{Data Availability}

Original FAST observational data will be open resources according to the FAST data 1-year protection policy. The folded and calibrated pulsar profiles presented in this paper can be found on the webpage: \url{http://zmtt.bao.ac.cn/psr-fast/}.

\vspace{5mm}
\bibliographystyle{raa}

\end{document}